\documentclass{article}

\usepackage[english]{babel}
\usepackage[letterpaper,top=2cm,bottom=2cm,left=3cm,right=3cm,marginparwidth=2cm]{geometry}
\usepackage[colorlinks=true, allcolors=blue]{hyperref}

\usepackage{graphicx}
\usepackage{amsmath}
\usepackage{amssymb}
\usepackage{bm}
\usepackage{url}
\usepackage{xr}

\usepackage{natbib}
\usepackage{listings}
\usepackage{float}
\usepackage{titlesec}
\usepackage{chngcntr}

\newcommand{\indep}{\perp \!\!\! \perp}

\title{Blinded sample size recalculation in randomized controlled trials with analysis of covariance}
\author{Takumi Kanata$^{1,2}$, Yasuhiro Hagiwara$^1$, and Koji Oba$^3$}
\date{%
    $^1$Department of Biostatistics, School of Public Health, Graduate School of Medicine, The University of Tokyo, Tokyo, Japan\\%
    $^2$Department of Clinical Data Science, Clinical Research \& Education Promotion Division, National Center of Neurology and Psychiatry, Tokyo, Japan\\%
    $^3$Interfaculty Initiative in Information Studies, The University of Tokyo, Tokyo, Japan%
}

\begin{document}

\maketitle

\begin{abstract}
In randomized controlled trials, covariate adjustment can improve statistical power and reduce the required sample size compared with unadjusted estimators. Analysis of covariance (ANCOVA) is often used to adjust for baseline covariates when outcomes are continuous. To design a sample size based on ANCOVA, it is necessary to prespecify the association between the outcome and baseline covariates, as well as among the baseline covariates themselves. However, determining these parameters at the design stage is challenging. Although it may be possible to adaptively assess these during the trial and recalculate the required sample size, existing sample size recalculation methods assume that the joint distribution of the outcome and baseline covariates is multivariate normal, which is not always the case in practice. In this study, we propose a blinded sample size recalculation method for the ANCOVA estimator based on the asymptotic relative efficiency under minimal distributional assumptions, thus accommodating arbitrary model misspecification. The proposed method is able to achieve the nominal power and reduce the required sample size without inflating the type I error rate in the final analysis. We conducted simulations to evaluate the performance of the proposed method under various scenarios. Applied to an HIV clinical trial, the proposed method reduced the required sample size by up to one-third compared to the unadjusted analysis, despite occasional slight increases.
\end{abstract}

\section{Introduction}\label{sec:intro}
In randomized controlled trials (RCTs), it is common not only to compare summary statistics of outcomes between treatment arms, but also to test and estimate treatment effects adjusted for baseline covariates that were measured prior to treatment assignment. Covariate adjustment can enhance the precision of treatment effect estimation, increase statistical power, and reduce the required sample size, thereby shortening the trial duration and lowering associated costs. Both the Food and Drug Administration \citep{FDA2023} and European Medicine Agency \citep{EMA2015} have issued guidance and generally recommend covariate adjustment in RCTs. When the outcomes are continuous, covariate adjustment is often performed using analysis of covariance (ANCOVA). If treatment assignment is independent of the baseline covariates, the ordinary least squares (OLS) estimator derived from the ANCOVA model is a consistent estimator of the treatment effect, regardless of model misspecification, and is known to be asymptotically at least as efficient as, or more efficient than, the unadjusted estimator based on the outcome mean in each treatment arm \citep{Yang2001}. 

However, at the design stage, it is often difficult to determine the extent to which covariate adjustment will improve efficiency and reduce the required sample size. When covariate adjustment is performed using ANCOVA for continuous outcomes, several methods have been proposed to calculate the required sample size, under the assumption that the joint distribution of the outcome and baseline covariates is multivariate normal. These methods require prespecification of the correlations between the outcome and baseline covariates \citep{Borm2007}, or conditional variances of the outcome given the treatment and baseline covariates \citep{Shieh2017}. However, it is often difficult to determine such correlation coefficients and conditional variances at the design stage. \cite{Li2023} proposed a general method to estimate the degree of efficiency gain from covariate adjustment at the design stage using external data. Their approach assumes that the conditional distribution of the outcome given the baseline covariates used for adjustment is identical between the external data and the control arm of the RCT. Nevertheless, this assumption may not hold when the planned RCT includes a placebo control and placebo effects are expected. Moreover, suitable external data may not be available. In summary, determining the potential reduction in the required sample size through covariate adjustment at the design stage is often challenging.

To address this issue, several methods have been proposed to recalculate the required sample size for ANCOVA through blinded interim analyses \citep{Friede2011, Zimmermann2020}. Because these analyses are conducted under blinding, they would not affect the type I error rate in the final analysis. Nevertheless, these methods assume that the joint distribution of the outcome and baseline covariates is multivariate normal. In practice, however, this assumption is often unrealistic, and discrete variables, such as sex, are frequently used in covariate adjustment. Under such conditions, the validity of the aforementioned sample size (re)calculation methods based on ANCOVA remains questionable. In general, a group sequential design known as an information-adaptive design \citep{Van-Lancker2025} has been proposed, in which the information level accounting for covariate adjustment is monitored sequentially, and an interim analysis is conducted when a prespecified information threshold is reached. When covariate adjustment improves efficiency, information accrual for the covariate-adjusted estimator is faster than that for the corresponding unadjusted estimator, resulting in a shorter trial duration. However, this approach imposes an operational burden due to the need for continuous monitoring.

We propose a blinded sample size recalculation method for RCTs with a $1:1$ allocation ratio where treatment is assigned independently of baseline covariates, and covariate adjustment is performed using ANCOVA for continuous outcomes. The proposed method does not require assumptions regarding the joint distribution of the outcome and baseline covariates, nor does it rely on correct specification of the ANCOVA model. In the proposed method, the initial sample size is determined based on an unadjusted estimator at the design stage, and sample size recalculation is conducted during an interim analysis. Consequently, no prior knowledge of the degree of efficiency gain from covariate adjustment is required. Furthermore, because the proposed method does not impose assumptions regarding the baseline covariate distribution, it is applicable even when adjusting for non-normally distributed or discrete covariates. It has the additional advantage that the interim analysis can be conducted under blinding, minimizing practical or operational biases that may affect clinical trial validity.

The remainder of this paper is organized as follows. In Section \ref{sec:methods}, we introduce the notation and setup and describe the proposed approach to recalculate sample size based on ANCOVA. In Section \ref{sec:application}, we apply the method to data from the AIDS Clinical Trials Group Protocol 175 (ACTG 175) \citep{Hammer1996}, and demonstrate the performance of the proposed method through extensive simulation studies in Section \ref{sec:simulation}. We end off with a discussion in Section \ref{sec:discussion}. 

\section{Methods}\label{sec:methods}

\subsection{Notation and setup}
We consider an RCT with total sample size $n$. For patient $i \in \{1, \cdots,n\}$, let $A_i$ be an indicator for treatment assignment ($1$ for the experimental treatment arm; $0$ for the control arm), $\bm{W}_i\in\mathbb{R}^k$ the $k$-dimensional baseline covariate vector observed prior to treatment assignment, and $Y_i$ a continuous outcome. The first and second moments of $(\bm{W}_i, Y_i)$ are assumed to exist. Data of patient $i$, $(A_i, \bm{W}_i, Y_i)$, are an independent and identically distributed sample drawn from an unknown distribution. We also assume that the treatment is randomly assigned with ratio $1:1$, independent of the baseline covariates; that is, 
\begin{align}
    P(A = 1) &= 0.5 \label{eq:equal allocation}
\end{align}
and
\begin{align}
    A \indep& \bm{W}. \label{eq:independent allocation}
\end{align}
Denote by $Y_i^a$ the potential outcome corresponding to $A_i=a$. The estimand of interest is the average causal effect: 
\begin{align*}
    \Delta = E[Y^1 - Y^0].
\end{align*}
Under the assumptions (\ref{eq:equal allocation}) and (\ref{eq:independent allocation}), and the consistency assumption $(Y_i=A_iY_i^1+(1-A_i)Y_i^0)$, $\Delta$ is identifiable from observed data $\{(A_i, \bm{W}_i, Y_i)\}_{i=1}^n$, that is, $\Delta = E[Y|A=1] - E[Y|A=0]$. 

We consider the following ANCOVA model without interaction terms between treatment assignment and baseline covariates: 
\begin{align}
    E[Y|A,\bm{W}] = \beta_0 + \beta_A A + \bm{\beta}_{\bm{W}}^\top \bm{W}. \label{eq:ancova}
\end{align}
For the unadjusted model, we also consider the following saturated model: 
\begin{align}
    E[Y|A] = \alpha_0 + \alpha_A A. \label{eq:unadjusted}
\end{align}
Let $\widehat{\beta}_0, \widehat{\beta}_A, \widehat{\bm{\beta}}_{\bm{W}}, \widehat{\alpha}_0,$ and $\widehat{\alpha}_A$ be OLS estimators in models (\ref{eq:ancova}) and (\ref{eq:unadjusted}), and $\underline{\beta}_0, \underline{\beta}_A, \underline{\bm{\beta}}_{\bm{W}}, \underline{\alpha}_0,$ and $\underline{\alpha}_A$ be the convergence in probability of each OLS estimator, respectively. Under the above assumptions, the OLS estimators are then consistent $(\Delta = \underline{\beta}_A = \underline{\alpha}_A)$ regardless of model misspecification \citep{Yang2001}. 

In what follows, we consider a one-sided test under the framework of a superiority trial, where the null hypothesis $H_0$ is that $\Delta \leq 0$, and the alternative hypothesis $H_1$ that $\Delta > 0$. However, the proposed method is also applicable to equivalence and non-inferiority trials. We prespecify the one-sided type I error rate and power to be $\alpha$ and $1 - \beta$, respectively. In the conventional sample size calculation based on the unadjusted model (\ref{eq:unadjusted}) under the assumption that the outcome variance is common across the treatment arms, we prespecify the treatment effect $\delta$ and the within-arm variance of the outcome $\sigma_Y^2$ based on the background knowledge, including evidence from prior trials \citep{Gould1995}, and calculate the total sample size as follows, 
\begin{align}
    N_{\text{unadj}} = \frac{4 (z_{1-\alpha} + z_{1-\beta})^2 \sigma_Y^2}{\delta^2}. \label{eq:n unadj}
\end{align}
where $z_p$ denotes the $p$-quantile of the standard normal distribution. In general, when we consider a balanced design (i.e., $P(A=1)=0.5$), the total sample size should be even. Thus, if the (re)calculated total sample size is odd, we add one hereafter. In this study, we assume that the sample size was calculated using formula (\ref{eq:n unadj}) at the design stage and recalculate it based on the ANCOVA model (\ref{eq:ancova}) at the time of interim analysis as $N_{\tau} = \tau N_{\text{unadj}}\ (0 < \tau < 1)$ under blinding.

\subsection{Blinded sample size recalculation based on ANCOVA}
Under the assumptions (\ref{eq:equal allocation}) and (\ref{eq:independent allocation}), the asymptotic relative efficiency of the ANCOVA estimator $\widehat{\beta}_A$ and the unadjusted estimator $\widehat{\alpha}_A$ is \citep{Wang2019}
\begin{align}
    \frac{AVar(\widehat{\beta}_A)}{AVar(\widehat{\alpha}_A)} = \frac{Var(Y - \underline{\beta}_A A - \underline{\bm{\beta}}_{\bm{W}}^\top \bm{W})}{Var(Y - \underline{\alpha}_A A)}, \label{eq:relative efficiency}
\end{align}
where $AVar$ denotes the asymptotic variance. Note that equation (\ref{eq:relative efficiency}) holds under arbitrary model misspecification of the ANCOVA model (\ref{eq:ancova}). In the numerator of equation (\ref{eq:relative efficiency}), it follows from assumptions (\ref{eq:equal allocation}) and (\ref{eq:independent allocation}) that
\begin{align}
    Var(Y - \underline{\beta}_A A - \underline{\bm{\beta}}_{\bm{W}}^\top \bm{W}) 
    = Var(Y-\underline{\beta}_0-\underline{\bm{\beta}}_{\bm{W}}^\top \bm{W}) - \frac{\Delta^2}{4} \label{eq:numerator}
\end{align}
(see Web Appendix \ref{suppsubsec:proof of numerator} for the proof). Consider, now, the OLS estimators $\widehat{\gamma}_0$ and $\widehat{\bm{\gamma}}_{\bm{W}}$ of the linear regression model $E[Y|\bm{W}] = \gamma_0 + \gamma_{\bm{W}}^\top \bm{W}$, with $\widehat{\bm{\gamma}}_{\bm{W}} \xrightarrow{P} \underline{\bm{\beta}}_{\bm{W}}$, following from assumption (\ref{eq:independent allocation}) (see Web Appendix \ref{suppsubsec:proof of convergence} for the proof). Since the true treatment effect $\Delta$ is unknown, we assume that the prespecified treatment effect is true (i.e., $\delta = \Delta$), and substitute $\delta$ for $\Delta$. Therefore, at the time of interim analysis $N_{\tau}$, using a consistent estimator of the residual variance of the linear regression model $E[Y|\bm{W}] = \gamma_0 + \bm{\gamma}_{\bm{W}}^\top\bm{W}$ and the prespecified treatment effect $\delta$, the consistent estimator of the numerator of equation (\ref{eq:relative efficiency}) is 
\begin{align}
    \frac{1}{N_{\tau} - 1 - k} \sum_{i=1}^{N_{\tau}} (Y_i-\widehat{\gamma}_0-\widehat{{\bm{\gamma}}}^\top_{\bm{W}}\bm{W}_i)^2
    - \frac{\delta^2}{4}. \label{eq:est of num}
\end{align}
Although $\widehat{\gamma}_0$ does not necessarily converge to $\underline{\beta}_0$ in probability, the estimator (\ref{eq:est of num}) remains consistent because variance is invariant under location shifts. In the same way, the consistent estimator of the denominator of equation (\ref{eq:relative efficiency}) is
\begin{align}
    \frac{1}{N_{\tau}-1} \sum_{i=1}^{N_{\tau}}(Y_i-\overline{Y})^2 - \frac{\delta^2}{4}, \label{eq:est of den}
\end{align}
where $\overline{Y} = \sum_{i=1}^{N_{\tau}}Y_i/N_{\tau}$. This is similar to the estimator of the within-arm variance of the outcome $\sigma_Y^2$ based on the pooled sample, $\widehat{\sigma}_Y^2 = \frac{N_{\tau}-1}{N_{\tau}-2} \times$(\ref{eq:est of den}) \citep{Gould1995}. As $N_{\tau} \to \infty$, we have that $\widehat{\sigma}_Y^2 \to$ (\ref{eq:est of den}). Equation (\ref{eq:est of num}) is an extension of this equation incorporating covariate adjustment.

Using the above results, we propose the following sample size recalculation formula.
\begin{align}
    \widehat{N}_{\text{rec}} = N_{\text{unadj}} \times 
    \frac{\sum_{i=1}^{N_{\tau}}(Y_i-\widehat{\gamma}_0-\widehat{\bm{\gamma}}^\top_{\bm{W}}\bm{W}_i)^2/(N_{\tau}-1-k)-\delta^2/4}
    {\min\{\sigma_Y^2, \sum_{i=1}^{N_{\tau}}(Y_i-\overline{Y})^2/(N_{\tau}-1)-\delta^2/4\}}
    + \frac{z_{1-\alpha}^2}{2}. \label{eq:prop}
\end{align}
The second term is for the exact correction when the two-sample $t$-test with pooled variance based on the ANCOVA estimator $\widehat{\beta}_A$ is used at the final analysis \citep{Guenther1981, Schouten1999}. This adjustment works especially well for a small sample, and related studies have also employed this correction \citep{Friede2011, Zimmermann2020}. The $\min(\cdot)$ function in the denominator of the first term prevents an excessive reduction in the recalculated sample size when the estimated pooled variance (\ref{eq:est of den}) overestimates the true variance. We propose the following sample size recalculation procedure:
\begin{enumerate}
    \item At the design stage, prespecify the treatment effect $\delta$ and the within-arm variance of the outcome $\sigma_Y^2$, and calculate the total sample size $N_{\text{unadj}}$ using formula (\ref{eq:n unadj}). Also prespecify the time of interim analysis $N_{\tau} = \tau N_{\text{unadj}}$ (e.g., $\tau = 0.5$). 
    \item At the interim analysis $N_{\tau}$, calculate $\widehat{N}_{\text{rec}}$ using formula (\ref{eq:prop}) under blinding. 
    \item Finally, specify the required sample size at the final analysis as follows:
    \begin{align*}
        \widehat{N}_{\text{fin}} = \min\{\max\{N_{\tau}, \widehat{N}_{\text{rec}}\}, N_{\max}\}
    \end{align*}
    where $N_{\max} = mN_{\text{unadj}}, m\geq 1$, is an upper bound of the available sample size. 
\end{enumerate}
The upper bound $N_{\max}$ is required because misspecification of the design-stage parameters ($\delta$ and $\sigma_Y^2$) may result in an excessively large recalculated sample size. In addition, the recalculated sample size may occasionally exceed the originally planned sample size when the baseline covariates are less prognostic for the outcome than anticipated, as the asymptotic efficiency gain from ANCOVA adjustment may not always be fully realized in finite samples. Since the recalculation procedure is conducted in a blinded manner without using a treatment effect estimate, inflation of the type I error rate is not expected \citep{Kieser2003}. Consistent with this expectation, our simulation studies demonstrated that the type I error rate was adequately controlled (Section \ref{sec:simulation}). The R and SAS macro codes for implementing the proposed method are provided in Web Appendix \ref{suppsec:code}. 

At the final analysis, a $t$-test is conducted. In practice, functions in standard statistical software (e.g., \texttt{summary.lm} in R and \texttt{proc reg} in SAS) compute the following model-based ANCOVA variance estimator,
\begin{align}
    \widehat{Var}(\widehat{\beta}_A) = \frac{\widehat{Var}(Y-\widehat{\beta}_0-\widehat{\beta}_A-\widehat{\bm{\beta}}_{\bm{W}}^\top\bm{W})}
    {(n-1)\{ \widehat{Var}(A) - \widehat{Cov}(\bm{W},A)^\top \widehat{Var}(\bm{W})^{-1}\widehat{Cov}(\bm{W},A) \}}, \label{eq:var}
\end{align}
where $\widehat{Var}$ and $\widehat{Cov}$ are the sample variance and covariance, respectively, with degrees of freedom taken into account. This variance estimator has been shown to be consistent under assumptions (\ref{eq:equal allocation}) and (\ref{eq:independent allocation}) and arbitrary model misspecification \citep{Wang2019}. Thus, the $t$-test using the estimated variance is valid at the final analysis. Confidence intervals are also calculated using the same estimated variance. 

In previous studies, \cite{Friede2011} and \cite{Zimmermann2020} proposed a blinded sample size recalculation method for ANCOVA (hereafter referred to as the simple recalculation method) in which they calculate the unbiased estimator of residual variance of the linear regression model $E[Y|\bm{W}] = \gamma_0 + \bm{\gamma}_{\bm{W}}^\top\bm{W}$ and replace $\sigma_Y^2$ in equation (\ref{eq:n unadj}) with the estimated residual variance and add the exact correction term $(z_{1-\alpha}^2/2)$, as follows:
\begin{align*}
    \frac{4(z_{1-\alpha}+z_{1-\beta})^2}{\delta^2} \times \frac{1}{N_{\tau}-1-k}
    \sum_{i=1}^{N_{\tau}}(Y_i-\widehat{\gamma}_0-\widehat{\bm{\gamma}}_{\bm{W}}^\top\bm{W}_i)^2
    + \frac{z_{1-\alpha}^2}{2}.
\end{align*}
The derivation of this formula assumes that the joint distribution of the outcome and baseline covariates is multivariate normal. In contrast, the proposed method is motivated by asymptotic relative efficiency under arbitrary model misspecification and minimal distributional assumptions.

\section{Application to ACTG 175 Trial}\label{sec:application}
As an application, we reanalyzed data from the ACTG 175 trial with the application of the proposed method. The dataset was downloaded from the \texttt{speff2trial} R package. In ACTG 175, 2139 HIV-infected participants were enrolled and randomly assigned to four different antiretroviral regimens in an equal ratio: zidovudine (ZDV) monotherapy, didanosine (ddl) monotherapy, ZDV+didanosine therapy, and ZDV+zalcitabine therapy. In this study, we considered two of the groups: ZDV monotherapy (control) and ZDV+ddl therapy (experimental treatment). We considered CD4 count (cells/mm$^3$) at 20 $\pm$ 5 weeks as the outcome of interest, and evaluated the treatment effect using a one-sided test with a type I error rate of $\alpha = 0.025$ and power of $1-\beta = 0.8$. For covariate adjustment, we considered the following baseline covariates: for continuous covariates, CD4 count (cells/mm$^3$), CD8 count (cells/mm$^3$), age (years), weight (kg), and Karnofsky score (on a scale of 0-100); and the following binary covariates: hemophilia, homosexual activity, history of intravenous drug use, race (White or Non-white), sex (female or male), antiretroviral history (naive or experienced), and symptomatic status. 

In the original dataset, 532 participants were assigned to the ZDV monotherapy group, and 522 to the ZDV+ddl therapy group. The unadjusted mean difference of CD4 count at 20 $\pm$ 5 weeks was 67.033 (95\% confidence interval, $49.617$--$84.449$; $p<0.001$), and the pooled outcome variance was $Var(Y)=(147.855)^2$. Although there was moderate heteroskedasticity in the outcome (with an outcome variance of $(130.962)^2$ in ZDV monotherapy and $(156.304)^2$ in ZDV+ddl therapy), we ignored it for simplicity and calculated a within-arm variance $Var(Y|A)=(147.855)^2-(67.033)^2\times(522/1054)\times(532/1054)=(144.006)^2$. Based on these results, the required sample size for the unadjusted estimator was calculated to be 146. Accordingly, we randomly drew 73 participants from each arm of the original dataset. The baseline characteristics of the selected participants are shown in Web Table~\ref{tab:baseline}, and sample correlations between CD4 count at 20 $\pm$ 5 weeks and the baseline covariates in the selected participants are shown in Web Table~\ref{tab:corr}. For covariate adjustment, we considered the following six strategies: adjustment for (1) CD4 count only, (2) antiretroviral history only, (3) CD4 count and antiretroviral history, (4) all continuous baseline covariates, (5) all binary baseline covariates, and (6) all baseline covariates. The reason for our planned adjustment for antiretroviral history only was that this factor was used as a stratified factor: to be precise, the duration of prior antiretroviral therapy was used. In the ANCOVA model, we included only a linear combination of the selected covariates, with neither higher-order nor interaction terms. We planned to conduct the interim analysis when the outcome had been observed for half of the required sample size ($\tau = 0.5$; that is, 73) and at this point, for each covariate adjustment strategy, apply the proposed method to recalculate the required sample size for the ANCOVA estimator. We set the upper bound of the available sample size to twice the required sample size for the unadjusted estimator $(m=2)$. If the required sample size exceeded 146, the remaining portion in each arm was randomly drawn from the original dataset, excluding participants who had already been selected. 

The results are summarized in Table~\ref{tab:application}. The one-sided null hypothesis was rejected in the all analyses, which suggested that ZDV+ddl therapy significantly improved CD4 count at 20 $\pm$ 5 weeks compared with ZDV monotherapy. These results are identical to those of the original analysis. However, the recalculated sample sizes varied. The minimum recalculated sample size was 100 when baseline CD4 count and antiretroviral history were adjusted, and the maximum recalculated sample size was 158, which exceeded the sample size needed for the unadjusted estimator $(N_{\text{unadj}}=146)$ when only antiretroviral history was adjusted for. When the baseline CD4 count was adjusted for, the recalculated sample size was small. Differences between the ANCOVA and unadjusted point estimates were primarily driven by the sampling process; because additional participants were drawn sequentially to meet the varying recalculated sample sizes, the final datasets differed across strategies (1)--(6), as well as from the dataset used for the unadjusted estimator. For example, the unadjusted estimate computed using the recalculated dataset for ANCOVA adjusting for (6) all baseline covariates was $88.521$, which differed from the unadjusted estimate obtained using the originally planned sample size. The ANCOVA estimator bias was evaluated using an extensive simulation study, as described in the following Section.

\begin{table}
\centering
\caption{Results of ACTG 175 trial reanalysis. Estimate of "'ZDV+ddl therapy" $-$ "ZDV monotherapy", standard error (SE), \textit{p}-value, 95\% confidence intervals (95\%CI), and (recalculated) sample size ((Recalculated) $N$). ANCOVA adjustment for (1) CD4 count only, (2) antiretroviral history only, (3) CD4 count and antiretroviral history, (4) all continuous baseline covariates, (5) all binary baseline covariates, and (6) all baseline covariates.}
\begin{tabular}{llrrrrr}
  \hline
  Estimator & & Estimate & SE & \textit{p}-value & 95\%CI & (Recalculated) $N$ \\ 
  \hline
  Unadusted &  & $72.562$ & $25.420$ & $0.002$ & $22.318$--$122.806$ & $146$ \\ 
  ANCOVA & (1) & $71.485$ & $24.158$ & $0.002$ & $23.633$--$119.337$ & $118$ \\ 
         & (2) & $75.011$ & $23.693$ & $0.001$ & $28.208$--$121.813$ & $158$ \\ 
         & (3) & $80.723$ & $24.480$ & $0.001$ & $32.131$--$129.315$ & $100$ \\ 
         & (4) & $74.001$ & $24.856$ & $0.002$ & $24.716$--$123.285$ & $112$ \\ 
         & (5) & $77.378$ & $23.653$ & $0.001$ & $30.623$--$124.133$ & $152$ \\ 
         & (6) & $82.273$ & $25.836$ & $0.001$ & $30.953$--$133.593$ & $104$ \\ 
   \hline
\end{tabular}
\label{tab:application}
\end{table}

\section{Simulation Study}\label{sec:simulation}

\subsection{Simulation scenarios}

We conducted simulations to evaluate the performance of the proposed method under three simple scenarios and a scenario that mimicked the ACTG 175 trial. In all scenarios, treatment $A$ was randomly assigned at a $1:1$ ratio independent of the baseline covariates, and we considered a one-sided test with a type I error rate of $\alpha = 0.025$ and power of $1-\beta=0.8$. In all three simple scenarios, the outcome $Y$ was assumed to follow a normal distribution with a within-arm variance $\sigma_Y^2=1$ in each group; five different effect sizes, $\Delta = \{0.3, 0.4, 0.5, 0.6, 0.7\}$, were considered. The dimensions of the baseline covariates were set to $k=2$, $(W_1,W_2)$, and the following three scenarios were considered:
\begin{itemize}
    \item Scenario (1): both $W_1$ and $W_2$ follow a standard normal distribution. Nine combinations were considered: three values of correlation between $W_1$ and $W_2$, $\sigma_{W_1W_2} = \{0.25, 0.5, 0.75\}$, and three values of correlation between $Y$ and each of $W_1$ and $W_2$, $\sigma_{Y\bm{W}} = \{0.25, 0.5, 0.75\}$. 
    \item Scenario (2): $W_1$ follows a binomial distribution, and $W_2$ a normal distribution. We generated $W_1,W_2$ and $Y$ from the following model: 
    \begin{align*}
        &W_1 \sim Ber(0.5), \quad W_2|W_1 \sim N(\mu_{W_2|W_1}^{(2)}W_1, 1), \\
        &E[Y|A,W_1,W_2] = \Delta A + \beta^{(2)}W_1 + \beta^{(2)}W_2,
    \end{align*}
    where a total of nine combinations were considered: three values of conditional expectation of $W_2$ given $W_1$, $\mu_{W_2|W_1}^{(2)} = \{0.5, 1, 1.5\}$, and three values of the outcome regression coefficient, $\beta^{(2)} = \{0.1, 0.3, 0.5\}$.
    \item Scenario (3): both $W_1$ and $W_2$ follow a binomial distribution. We generated $W_1,W_2$ and $Y$ from the following model: 
    \begin{align*}
        &W_1 \sim Ber(0.5), \quad W_2|W_1 \sim Ber(0.5 + \mu_{W_2|W_1}^{(3)}(W_1-0.5)), \\
        &E[Y|A,W_1,W_2] = \Delta A + \beta^{(3)}W_1 + \beta^{(3)}W_2,
    \end{align*}
    where a total of nine combinations are considered: three values of relationship between $W_1$ and $W_2$, $\mu_{W_2|W_1}^{(3)} = \{0.2, 0.4, 0.6\}$, and three values of the outcome regression coefficient, $\beta^{(3)} = \{0.1, 0.3, 0.5\}$.
\end{itemize}
Under these conditions, the required total sample sizes corresponding to the unadjusted estimator for each effect size were $N_{\text{unadj}} = \{350, 198, 126, 88, 66\}$. 

In scenario (4), we simulated data that mimic the ACTG 175 trial. We generated the baseline covariates specified in Section \ref{sec:application} using a latent variable approach designed to approximately preserve the dependence structure observed in the original data. Specifically, latent variables with mean zero were generated from a multivariate normal distribution whose correlation matrix was estimated from the original dataset. Continuous baseline covariates were obtained by appropriately shifting and scaling the latent variables, whereas binary baseline covariates were generated by thresholding latent variables to match the observed marginal probabilities. For the outcome, we obtained, via regression, CD4 counts at 20 $\pm$ 5 weeks on all of the baseline covariates in each arm separately, and calculated its conditional mean. The conditional variances $Var(Y|A=1,\bm{W})$ and $Var(Y|A=0,\bm{W})$ were also calculated based on the squared residuals, although we set the within-arm variances $Var(Y|A=1)$ and $Var(Y|A=0)$ to be identical. The resulting true treatment effect was $70.303$, and the true within-arm variance $(143.615)^2$. Thus, the required sample size corresponding to the unadjusted estimator was 132. The details are presented in Web Appendix \ref{suppsec:scenario4}.

In all of the scenarios, we assumed that we correctly prespecified the true effect size and the within-arm variance at the design stage, and the interim analysis was conducted at $\tau = 0.5$. We set the upper bound of the available sample size to twice the required sample size for the unadjusted estimator $(m=2)$. The required sample size $\widehat{N}_{\text{fin}}$ based on the ANCOVA model (\ref{eq:ancova}) was recalculated using the proposed method, as well as the simple recalculation method for comparison. In scenario (1), the exact sample size was also calculated by the method of \cite{Shieh2017}, and we varied the value of $\tau$ from 0.1 to 0.9 in increments of 0.1 to evaluate the impact of the point in time of the interim analysis for $(\Delta, \sigma_{W_1W_2}, \sigma_{Y\bm{W}}) = (0.5, 0.5, 0.5)$. The type I error rate was also calculated for scenario (1) by setting $\Delta=0$ for every combination $(\sigma_{W_1W_2}, \sigma_{Y\bm{W}})$, maintaining $\delta = \{0.3,0.4,0.5,0.6,0.7\}$. Additionally, in scenario (4), we studied the performance of the proposed method when we over- and under-estimated the within-arm variance at the design stage (1.1 and 0.9 times the true variance, respectively).

For each scenario and setting, we conducted 100,000 Monte Carlo runs; calculated the Monte Carlo bias (and Monte Carlo percent bias for scenario (4)), empirical standard error, Monte Carlo power, and 95\% coverage; and subsequently summarized the recalculated sample sizes. 

\subsection{Simulation results}
\begin{figure}
    \centering
    \includegraphics[width=1\linewidth]{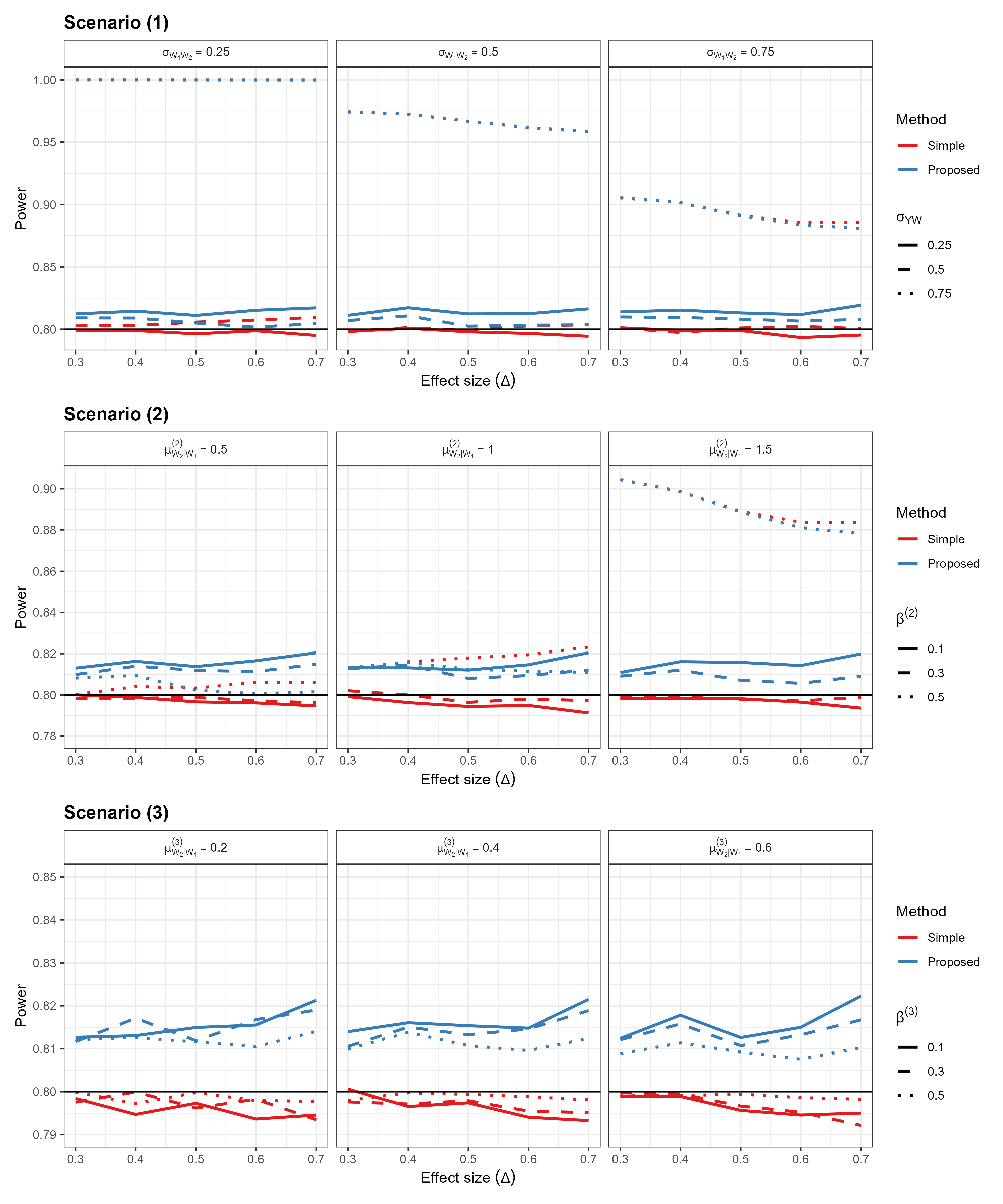}
    \caption{Monte Carlo powers of the simple recalculation method (Simple) and the proposed method (Proposed) for scenario (1)--(3).}
    \label{fig:scenario1-3 power}
\end{figure}

\begin{figure}[H]
    \centering
    \includegraphics[width=1\linewidth]{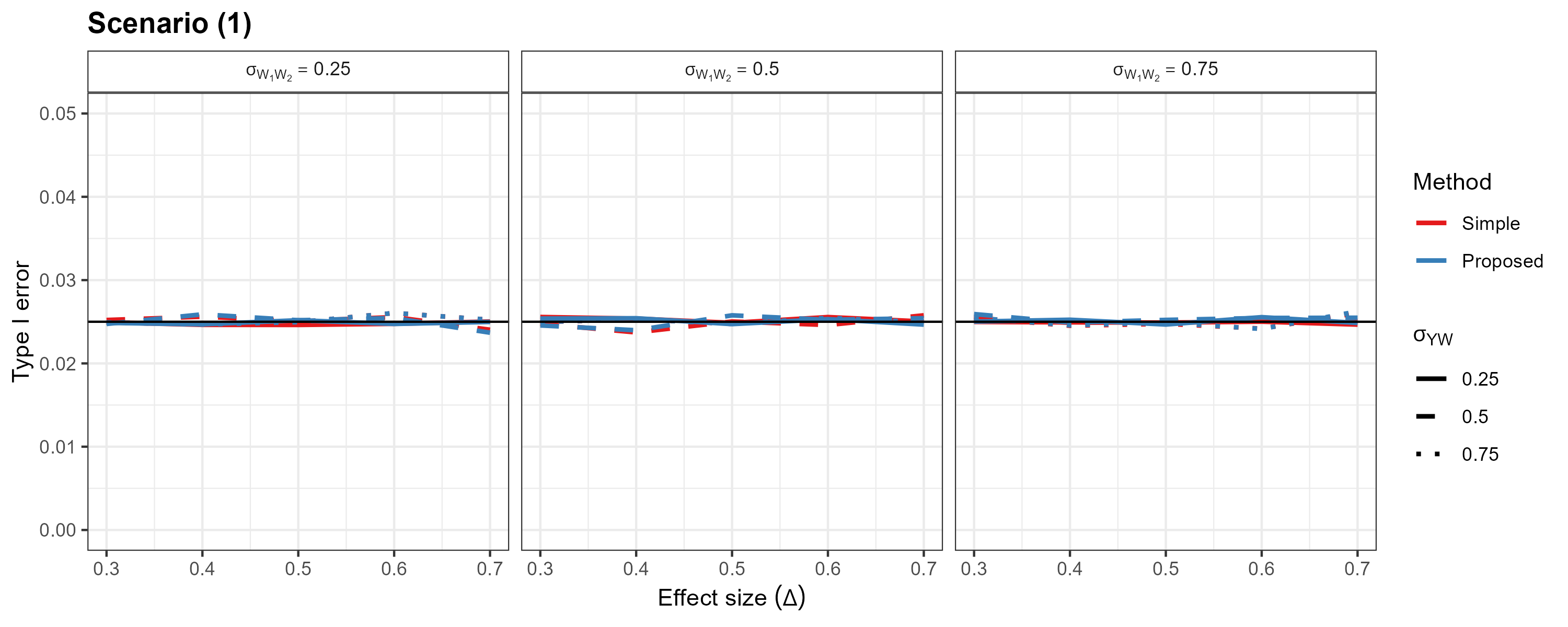}
    \caption{Monte Carlo type I error rate of the simple recalculation method (Simple) and the proposed method (Proposed) for scenario (1).}
    \label{fig:alpha}
\end{figure}

The powers of the simple recalculation method and the proposed method are shown in Figure~\ref{fig:scenario1-3 power}. Both the simple recalculation method and the proposed method achieved the nominal power of 80\%, although the proposed method was slightly conservative, especially when the covariates did not follow a normal distribution. This may be due to the $\min(\cdot)$ function in equation (\ref{eq:prop}). The powers of both the simple recalculation method and the proposed method greatly exceeded the nominal power when $\sigma_{Y\bm{W}} = 0.75$ in scenario (1), and when $(\mu_{W_2|W_1}^{(2)}, \beta^{(2)}) = (1.5, 0.5)$ in scenario (2), but this would be due to the required sample size for the ANCOVA being smaller than the accumulated sample size at the interim analysis. Neither the simple recalculation method nor the proposed method inflated the type I error rate in the final analysis (see Web Figure~\ref{fig:alpha}). Additional simulations demonstrate similar results under model misspecification (Web Appendix \ref{suppsubsubsec:misspecified}). Bias and 95\% coverage yielded acceptable results in scenarios (1)--(3) (see Web Tables~\ref{tab:scenario1}--\ref{tab:scenario3}).

\begin{figure}[H]
    \centering
    \includegraphics[width=1\linewidth]{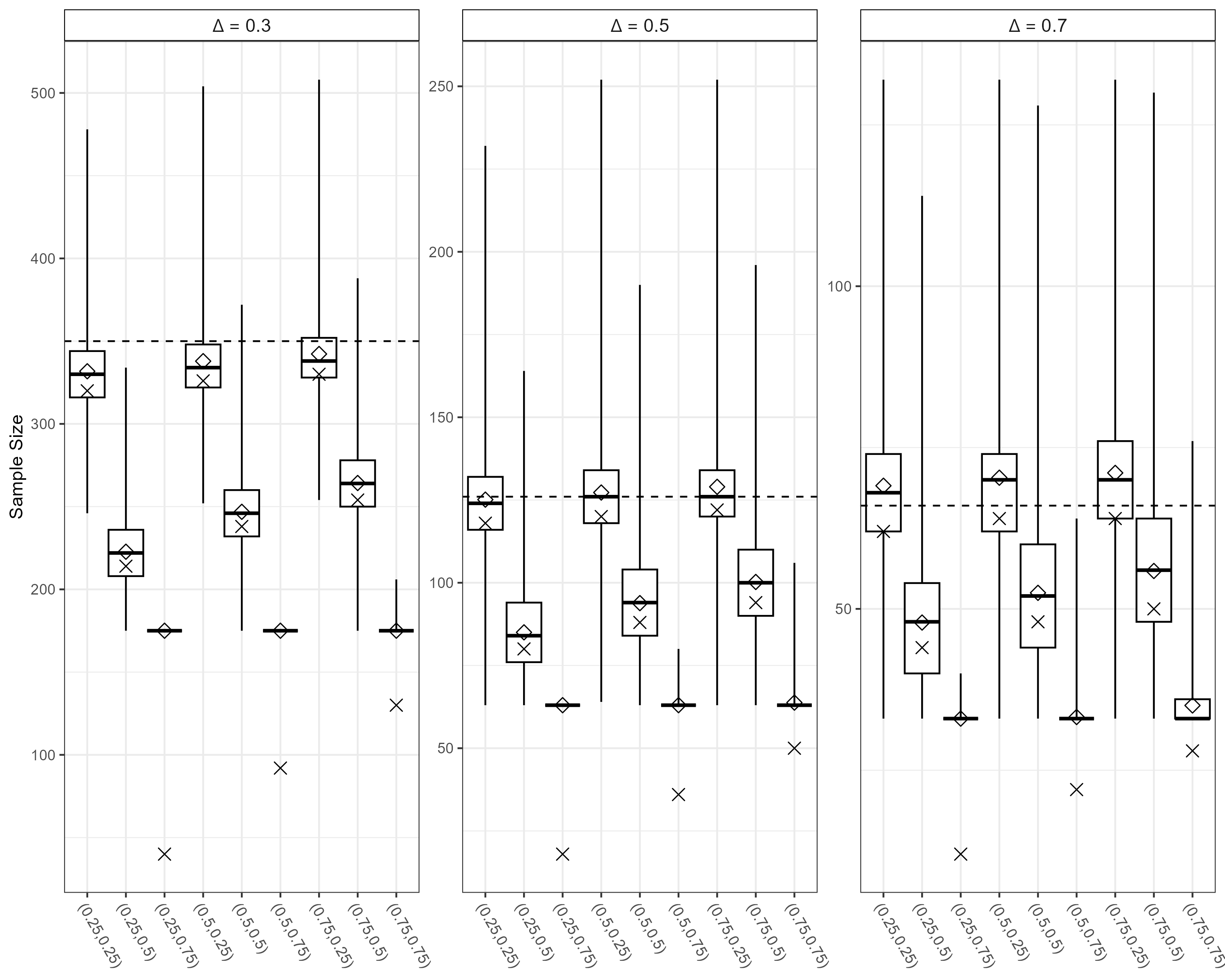}
    \caption{Recalculated sample sizes using the proposed method in scenario (1). Dashed horizontal lines indicate required sample sizes for the unadjusted estimator. Cross marks indicate exact sample sizes calculated according to \cite{Shieh2017}. Tick marks on the x-axis represent $(\sigma_{W_1W_2},\sigma_{Y\bm{W}})$.}
    \label{fig:scenario1 sample size}
\end{figure}

\begin{figure}[H]
    \centering
    \includegraphics[width=1\linewidth]{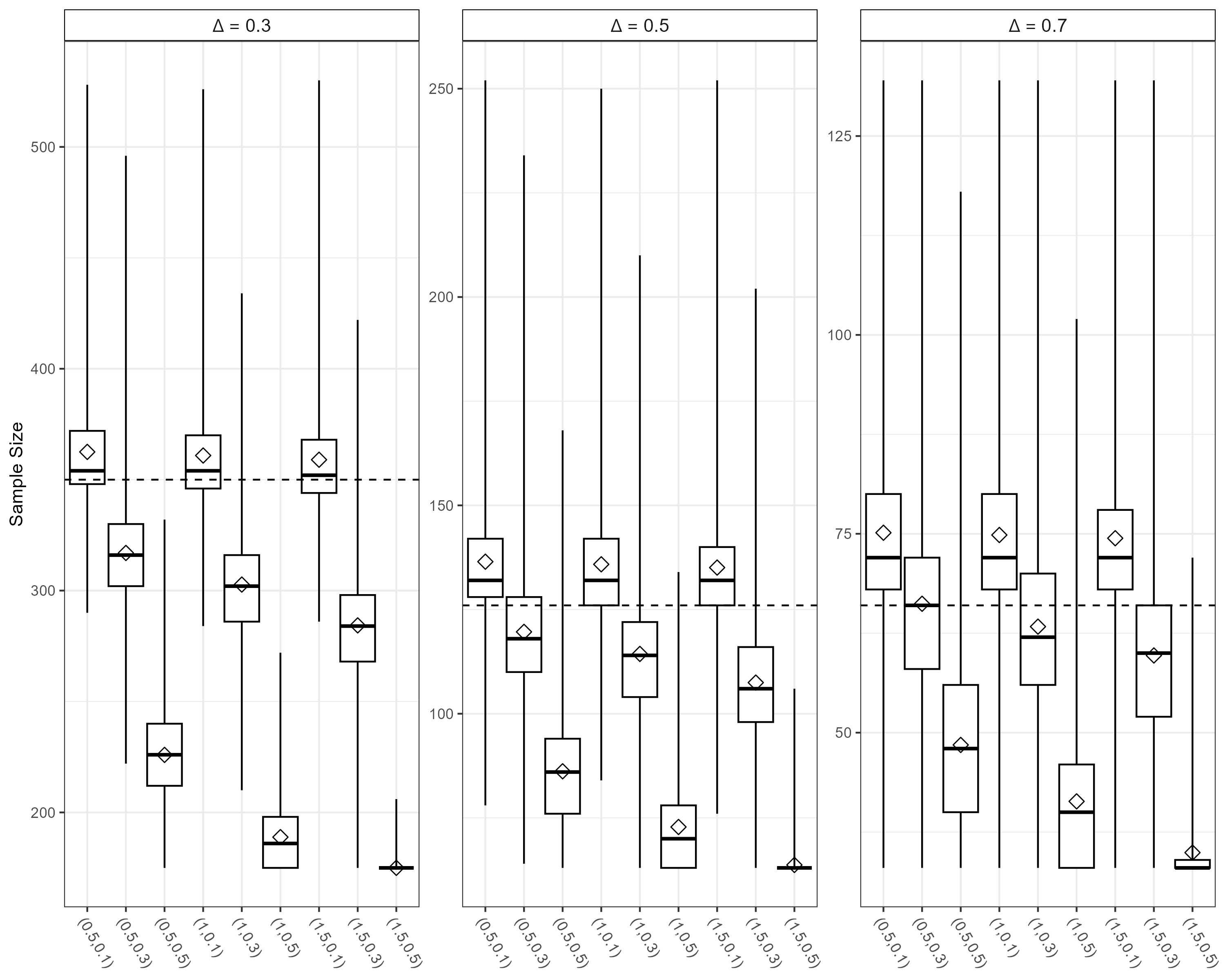}
    \caption{Recalculated sample sizes using the proposed method in scenario (2). Dashed horizontal lines indicate required sample sizes for the unadjusted estimator. Tick marks on the x-axis represent $(\mu_{W_2|W_1}^{(2)},\beta^{(2)})$.}
    \label{fig:scenario2 sample size}
\end{figure}

\begin{figure}[H]
    \centering
    \includegraphics[width=1\linewidth]{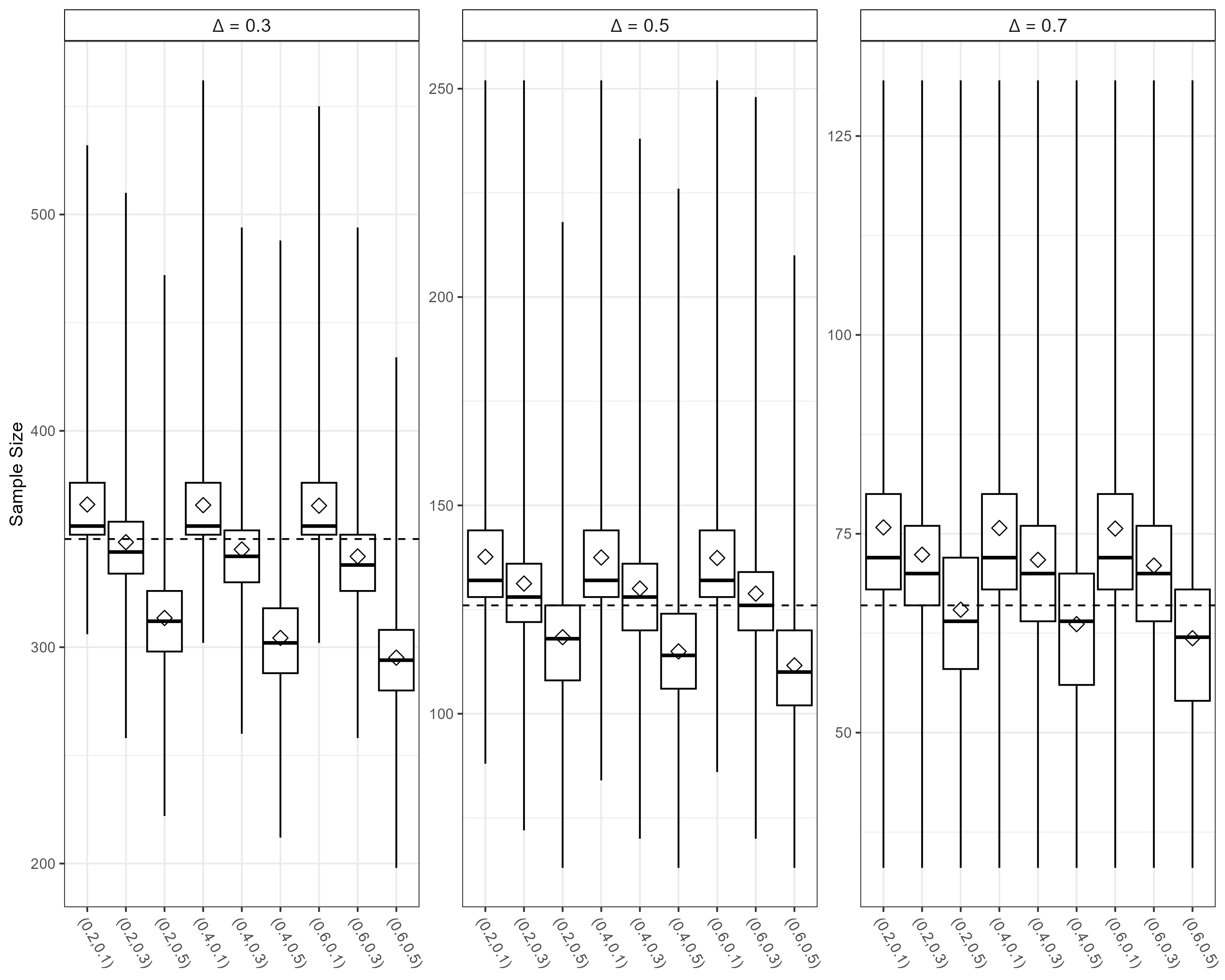}
    \caption{Recalculated sample sizes using the proposed method in scenario (3). Dashed horizontal lines indicate required sample sizes for the unadjusted estimator. Tick marks on the x-axis represent $(\mu_{W_2|W_1}^{(3)},\beta^{(3)})$.}
    \label{fig:scenario3 sample size}
\end{figure}

\begin{figure}[H]
    \centering
    \includegraphics[width=1\linewidth]{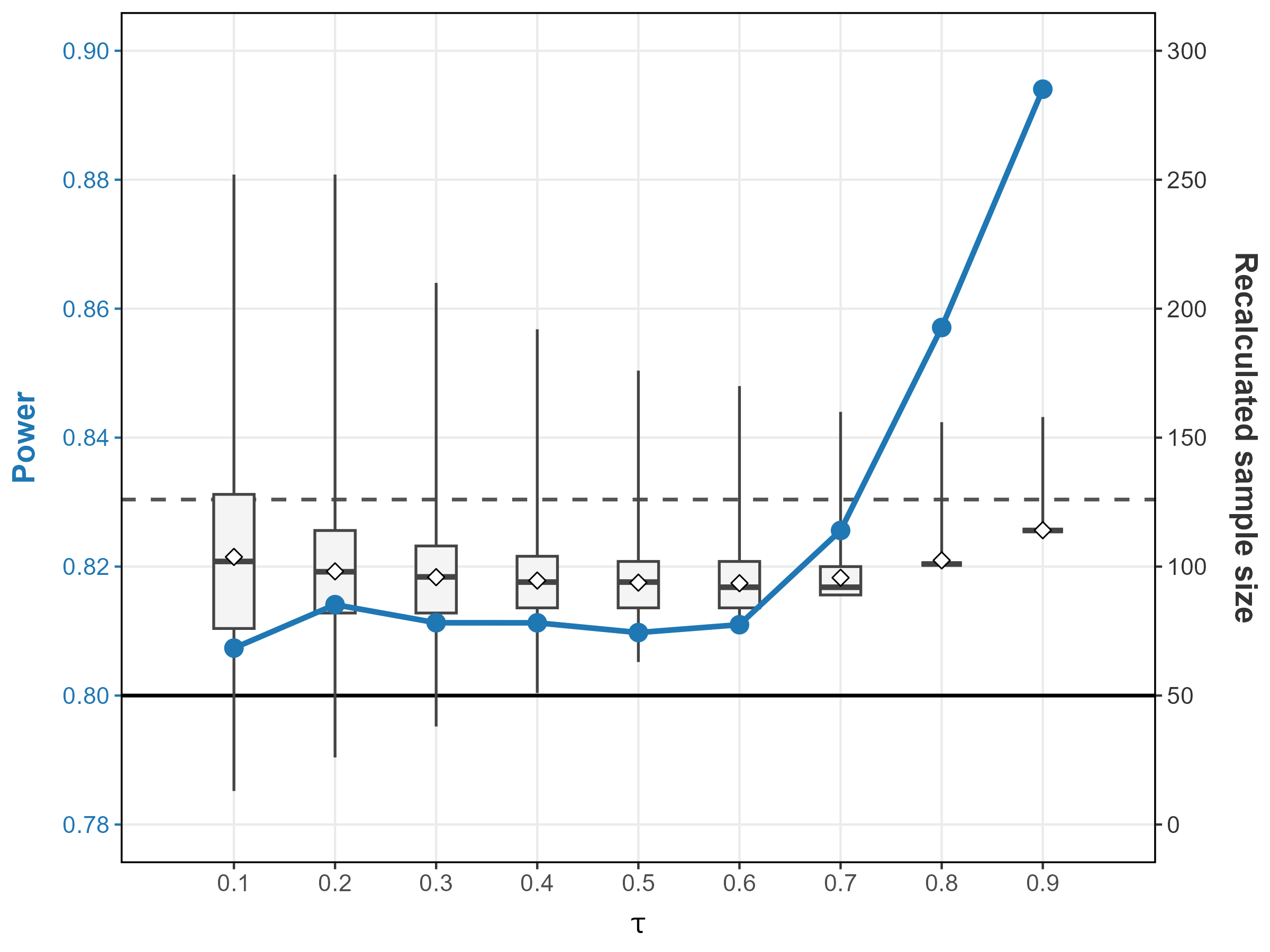}
    \caption{Monte Carlo power (blue line) and recalculated sample size by the proposed method (boxplot) for scenario (1), $(\Delta, \sigma_{W_1W_2}, \sigma_{Y\bm{W}}) = (0.5, 0.5, 0.5)$, varying the value of $\tau$ for the point in time when the interim analysis was conducted $N_{\tau} = \tau N_{\text{unadj}}$. Dashed horizontal line indicates required sample size for unadjusted estimator (126).}
    \label{fig:timing}
\end{figure}

Boxplots of the recalculated sample size based on the proposed method for $\Delta = \{0.3,0.5,0.7\}$ are shown in Web Figures~\ref{fig:scenario1 sample size}--\ref{fig:scenario3 sample size}. In addition, the exact sample size for the ANCOVA based on the method of \cite{Shieh2017} was also indicated for scenario (1). Observe that, in this scenario, the recalculated sample size based on the proposed method was very close to the exact sample size, and for $\sigma_{Y\bm{W}} = 0.75$, the exact sample size was smaller than the accumulated sample size at the interim analysis. The extent to which the sample size was reduced through covariate adjustment depended on the strength of the association between covariates and the outcome, as well as the strength of the associations among the covariates themselves. The reduction was primarily influenced by the strength of the association between the covariates and the outcome: the stronger the association, the smaller the recalculated sample size. However, when the association between the covariates and the outcome is weak, the recalculated sample size may exceed the required sample size for the unadjusted estimator, even though all parameters are correctly specified. The strength of the associations among the baseline covariates also contributed to a reduction in the required sample size, although the extent of the reduction was less than that due to the association with the outcome. When covariates followed a multivariate normal distribution (scenario (1)), a larger covariance between covariates contributed to a larger recalculated sample size. In contrast, both in scenarios (2) and (3), the recalculated sample size decreased when the association between the baseline covariates became stronger. The point in time when the interim analysis was conducted had little impact on the power, but an interim analysis at a later point in time provided a more accurate recalculation result. However, conducting the interim analysis too late did not lead to a sufficient reduction in the final sample size (see Web Figure~\ref{fig:timing}).

The results for scenario (4) are listed in Table~\ref{tab:scenario4}. Both the simple recalculation method and the proposed method yielded acceptable bias and coverage and achieved the nominal power of 80\%, especially when only a few baseline covariates $(k=1,2)$ were adjusted for. However, the simple recalculation method did not achieve the nominal power when all continuous $(k=5)$ or binary $(k=7)$ baseline covariates were adjusted for, whereas the proposed method achieved it. When all baseline covariates $(k=12)$ were adjusted for, both the simple recalculation method and proposed method were underpowered. Although the proposed method was generally conservative compared with the simple recalculation method, the recalculated sample sizes were nearly the same. Outcome data were simulated using a separate model for each arm; thus, the ANCOVA model was misspecified because it did not consider any interaction terms. However, as expected from the theoretical results, the proposed method performed well. Both the simple recalculation and proposed method showed slight negative bias. This is an inherent property of the sample size recalculation, as discussed in Section \ref{sec:discussion}.

The results for scenario (4) for the case where we misspecified the within-arm variance of the outcome at the design stage are shown in Web Table~\ref{tab:under over}. When the true variance was underestimated at the design stage (0.9 times the true variance), adjustment for a single covariate maintained a power of at least 0.79. Even when adjusting for all continuous or binary variables, a power of at least 0.77 was achieved. However, when all variables (including both continuous and binary variables) were adjusted for, the power decreased to 0.726. When we overestimated the true variance at the design stage (1.1 times the true variance), the proposed method achieved the nominal power of 80\%, except for the case of adjusting for all covariates, which yielded a power of 0.783.

In summary, both the simple recalculation method and the proposed method showed broadly comparable performance across most simulation scenarios. In scenario (4), however, the proposed method maintained the nominal power when adjusting for a larger number of baseline covariates, whereas the simple recalculation method did not.

\begin{table}
\centering
\small
\caption{Results for scenario (4) from 100,000 Monte Carlo runs. Monte Carlo bias (Bias), Monte Carlo percent bias (\%Bias), empirical standard error (EmpSE), Monte Carlo power (Power), 95\% coverage (Cov.), and summary of recalculated sample size ($N_\mathrm{Avg}$: mean, $N_\mathrm{Med}$: median, $N_\mathrm{Min}$: minimum, and $N_\mathrm{Max}$: maximum) for ANCOVA estimator, adjusting for (1) CD4 count only, (2) antiretroviral history only, (3) CD4 count and antiretroviral history, (4) all continuous baseline covariates, (5) all binary baseline covariates, and (6) all baseline covariates. "Simple" and "Prop." indicate the simple recalculation method and the proposed method, respectively.}
\begin{tabular}{lllrrrrrrrrr}
  \hline
  Estimator & & & Bias & \%Bias & EmpSE & Power & Cov & $N_\mathrm{Avg}$ & $N_\mathrm{Med}$ & $N_\mathrm{Min}$ & $N_\mathrm{Max}$ \\ 
  \hline
  Unadjusted &  &  & $-0.010$ & $-0.015$ & $25.099$ & $0.793$ & $0.949$ & \multicolumn{4}{c}{$132$} \\ 
  ANCOVA & (1) & Simple & $-1.318$ & $-1.874$ & $24.023$ & $0.805$ & $0.949$ & $99$ & $98$ & $66$ & $192$ \\ 
   &  & Prop. & $-0.898$ & $-1.277$ & $24.011$ & $0.812$ & $0.949$ & $98$ & $98$ & $66$ & $186$ \\ 
   & (2) & Simple & $-1.017$ & $-1.447$ & $24.394$ & $0.800$ & $0.949$ & $135$ & $134$ & $66$ & $260$ \\ 
   &  & Prop. & $-0.540$ & $-0.768$ & $24.024$ & $0.817$ & $0.950$ & $138$ & $134$ & $80$ & $254$ \\ 
   & (3) & Simple & $-1.334$ & $-1.897$ & $24.073$ & $0.802$ & $0.948$ & $95$ & $94$ & $66$ & $182$ \\ 
   &  & Prop. & $-0.953$ & $-1.356$ & $24.130$ & $0.806$ & $0.949$ & $95$ & $94$ & $66$ & $174$ \\ 
   & (4) & Simple & $-1.442$ & $-2.051$ & $24.559$ & $0.787$ & $0.949$ & $98$ & $96$ & $66$ & $190$ \\ 
   &  & Prop. & $-1.060$ & $-1.508$ & $24.564$ & $0.792$ & $0.949$ & $97$ & $96$ & $66$ & $184$ \\ 
   & (5) & Simple & $-1.381$ & $-1.964$ & $25.111$ & $0.776$ & $0.948$ & $129$ & $128$ & $66$ & $264$ \\ 
   &  & Prop. & $-0.872$ & $-1.240$ & $24.709$ & $0.793$ & $0.949$ & $131$ & $130$ & $66$ & $262$ \\ 
   & (6) & Simple & $-1.934$ & $-2.750$ & $25.665$ & $0.750$ & $0.948$ & $92$ & $90$ & $66$ & $176$ \\ 
   &  & Prop. & $-1.616$ & $-2.298$ & $25.769$ & $0.750$ & $0.948$ & $92$ & $90$ & $66$ & $170$ \\ 
   \hline
\end{tabular}
\label{tab:scenario4}
\end{table}

\section{Discussion}\label{sec:discussion}
In this study, we proposed a blinded sample size recalculation method for the ANCOVA estimator, based on the asymptotic relative efficiency of the ANCOVA and unadjusted estimators. The proposed method does not require unblinding, which would prevent the inflation of type I error rates, and is ideal for practical reasons. ICH E9 \citep{ICHE9} discussed sample size recalculations under blinding. If the sample size is recalculated based on unblinded data at the interim analysis, then the treatment effect may be estimable and could introduce bias in the trial. For example, investigators who know that the experimental treatment is highly beneficial could easily recognize participants in the control arm even in a double-blind trial and treat them carefully to compensate them for taking non-beneficial medications compared with the experimental treatment \citep{Proschan2005}. The proposed method is applicable under blinding, thereby minimizing the number of individuals who become aware of the treatment effect during the course of the trial, and reducing the potential for bias to the greatest extent possible. 

In our simulation study, the simple recalculation method \citep{Friede2011,Zimmermann2020} showed performance broadly comparable to that of the proposed method in most scenarios. This would suggest that the simple recalculation method may remain approximately valid beyond the settings under which it was originally derived, although its derivation relies on the assumption that the joint distribution of the outcome and baseline covariates is multivariate normal. We emphasize that the present work does not establish theoretical superiority of the proposed method over the existing simple recalculation method. Instead, the proposed method is motivated by the asymptotic relative efficiency under arbitrary model misspecification and minimal distributional assumptions. In practice, analyses adjusting for discrete covariates, including sex and stratification factors, are often conducted, and the proposed method is applicable in such settings. Moreover, in some scenarios, the proposed method achieved slightly higher power than the simple recalculation method while often requiring a smaller final sample size, especially when the baseline covariates were strongly prognostic of the outcome. The proposed method works well when a moderate number of covariates (e.g., $k=5-7$) are adjusted for; however, as previously reported \citep{Zimmermann2020}, the simple recalculation method could not achieve the nominal power even if three covariates were adjusted for. However, when we adjusted for many covariates (e.g., $k \geq 10$), the proposed method was unable to achieve the nominal power, similar to the simple recalculation method. The reason would be that the asymptotic approximation did not work well, and we confirmed that the proposed method achieved the nominal power in a large sample size, even when adjusting for more than ten covariates (Web Appendix \ref{suppsubsubsec:large-sample-size}). However, in real RCTs, it is uncommon to adjust for more than ten covariates, and adjusting for a smaller number is expected to yield sufficient efficiency gain. When the within-arm variance of the outcome was underestimated at the design stage, the proposed method was applicable to adjust for a few covariates. When we overestimated it, the power exceeded the nominal power, except when adjusting for too many covariates. 

In the simulations, both the simple recalculation and proposed methods tended to show slight negative bias, which may arise from the adaptive nature of the sample size recalculation procedure, where the final sample size is determined based on interim analysis data. Specifically, when the interim sample happens to exhibit a smaller treatment effect than the true value by chance, the pooled variance tends to be underestimated, leading the recalculation procedure to recommend a smaller final sample size and thus earlier trial termination. Conversely, when the interim sample exhibits a larger treatment effect, the pooled variance tends to be larger and the recalculated sample size tends to increase. In such cases, the additional observations accumulated after the interim analysis allow the estimated treatment effect to move closer to the true value through regression to the mean. This asymmetry may induce slight bias. Similar phenomena have been discussed in adaptive clinical trials \citep{Robertson2023}, including sample size recalculation for unadjusted estimators \citep{Denne2000}. Although we confirmed that the bias was negligible in large samples for the proposed method (Web Appendix \ref{suppsubsubsec:large-sample-size}), development of bias-reduced estimators under sample size recalculation remains a topic for future research.

However, the proposed method has several limitations. First, it only uses the observed outcomes, not the data of so-called pipeline participants, who are registered in the trial but whose outcomes are not observed at the time of interim analysis. Although conducting the interim analysis at a later time point would provide an accurate recalculation result, it is possible for the recruitment to complete at the time of the interim analysis, and the proposed method would not lead to a reduction in trial duration. Therefore, while the optimal timing depends on the specific context, our simulation results suggest that conducting interim analysis when outcomes have been observed for approximately half of the required sample size for the unadjusted estimator (i.e., $\tau=0.5$) provides a reasonable balance between recalculation accuracy and practical feasibility. If outcomes are collected longitudinally, one potential solution is to utilize short-term outcomes for sample size recalculation \citep{Van-Lancker2020}. Second, we assumed that the treatments were randomly assigned at a $1:1$ ratio, independent of the baseline covariates. If the allocation ratio is not $1:1$, equation (\ref{eq:relative efficiency}) is invalid and it would be difficult to apply the proposed method. Moreover, the model-based ANCOVA variance estimator (\ref{eq:var}) is generally inconsistent under arbitrary misspecification when randomization is unequal \citep{Bartlett2020}. Furthermore, in real clinical trial settings, covariate-adaptive randomization (CAR) methods, such as stratified randomization or minimization, are often used to balance important known risk factors. Additional simulations demonstrated similar empirical results under stratified randomization (Web Appendix \ref{suppsubsubsec:car}), although deriving the theoretical properties of the proposed method under CAR is challenging because CAR induces a non-i.i.d. data structure. \citet{Li2021} proposed a sample size recalculation method for linear models under CAR; however, their approach relies on restrictive assumptions, including correct specification of the linear model and independence among the stratification factors. Development of sample size (re)calculation procedures with theoretical guarantees under CAR is left for future investigation. Third, we only considered covariate adjustment for the continuous outcome; thus, our proposed method cannot be directly applied to the binary or time-to-event outcome. Covariate adjustment is conducted when the outcome is binary or time-to-event; for this case, the targeted maximum likelihood estimator (TMLE), which is based on semiparametric theory, is frequently used \citep{Benkeser2021}. Although the relative efficiency of the TMLE compared with the unadjusted estimator has also been discussed \citep{Moore2009,Moore2011}, its calculation based on pooled data would be difficult. Sample size (re)calculation for the covariate adjustment of the binary and time-to-event outcome is a topic for future research. Fourth, while this study focuses on fully observed data, missing outcomes are common in clinical trials. Although inverse probability weighting is a standard approach for handling missing outcome data, the efficiency gain obtained by combining weighting with regression adjustment methods such as ANCOVA is generally guaranteed only under restrictive assumptions \citep{Zhao2024}. Extending sample size (re)calculation procedures to settings with missing outcomes remains an important topic for future research. Finally, it should be noted that although covariate adjustment often leads to a reduction in the required sample size, decreasing the sample size may not always be appropriate. In particular, for safety outcomes, a relatively large sample size is required for observation. In addition, the ANCOVA estimator may be biased in small samples \citep{Tackney2023}, because it does not have the unbiasedness property under model misspecification. Therefore, we recommend that the final sample size should be determined not solely based on the proposed method, but through a comprehensive assessment that also considers safety and other relevant factors. 

\section*{Acknowledgment}
We would like to thank Dr. Yutaka Matsuyama for his helpful comments, and Editage (\url{www.editage.jp}) for English language editing.

\section*{Funding}
This work was supported by ”Work-Life Balance in Academia: A Career Skills Course for Researchers” Grant, Office for Gender Equity, Center for Coproduction of Inclusion, Diversity and Equity, The University of Tokyo.

\section*{Conflict of interest}
The authors declare no potential conflict of interest.

\section*{Data Availability}
The AIDS Clinical Trials Group Protocol 175 (ACTG 175) trial data used in Section \ref{sec:application} can be downloaded from the \texttt{speff2trial} R package (\url{https://cran.r-project.org/web/packages/speff2trial/index.html}).

\bibliographystyle{biom}
\bibliography{LaTexlec}

\newpage
\appendix

\renewcommand{\thesection}{\Alph{section}}
\renewcommand{\thesubsection}{\thesection.\arabic{subsection}}
\renewcommand{\tablename}{Web Table}
\renewcommand{\figurename}{Web Figure}

\section{Proofs}\label{suppsec:proof}
\subsection{Proof of equation (\ref{eq:numerator})}\label{suppsubsec:proof of numerator}
Observe that
\begin{align*}
    Var(Y-\underline{\beta}_AA-\underline{\bm{\beta}}_{\bm{W}}^\top\bm{W})
    &= Var(Y-\underline{\bm{\beta}}_{\bm{W}}^\top\bm{W}) + \underline{\beta}_A^2Var(A) 
    - 2Cov(Y-\underline{\bm{\beta}}_{\bm{W}}^\top\bm{W}, \underline{\beta}_AA) \\
    &= Var(Y-\underline{\bm{\beta}}_{\bm{W}}^\top\bm{W}) + \underline{\beta}_A^2Var(A) - 2\underline{\beta}_ACov(Y,A) \\
    &= Var(Y-\underline{\bm{\beta}}_{\bm{W}}^\top\bm{W}) + \underline{\beta}_A^2Var(A) - 2\underline{\beta}_A^2Var(A) \\
    &= Var(Y-\underline{\beta}_0-\underline{\bm{\beta}}_{\bm{W}}^\top\bm{W}) - \underline{\beta}_A^2Var(A) \\
    &= Var(Y-\underline{\beta}_0-\underline{\bm{\beta}}_{\bm{W}}^\top\bm{W}) - \frac{\Delta^2}{4},
\end{align*}
where the second equality follows from assumption (\ref{eq:independent allocation}), the third from $\underline{\beta}_A = Cov(Y,A)/Var(A)$, and the last from $\underline{\beta}_A=\Delta$ and assumption (\ref{eq:equal allocation}).

\newpage

\subsection{Proof of \texorpdfstring{$\widehat{\bm{\gamma}}_{\bm{W}} \xrightarrow{P} \underline{\bm{\beta}}_{\bm{W}}$}{TEXT}}\label{suppsubsec:proof of convergence}
Denote by $\bm{X} = (1, A, \bm{W}^\top)^\top$ the design matrix of the ANCOVA model (\ref{eq:ancova}). By using Slutsky's theorem, we obtain
\begin{align*}
    \begin{pmatrix}
        \widehat{\beta}_0 \\
        \widehat{\beta}_A \\
        \widehat{\bm{\beta}}_{\bm{W}}
    \end{pmatrix}
    = \left( n^{-1}\sum_{i=1}^n\bm{X}_i\bm{X}_i^\top \right)^{-1} \left( n^{-1}\sum_{i=1}^n \bm{X}_iY_i \right)
    \xrightarrow{P} \{ E[\bm{X}\bm{X}^\top] \}^{-1} E[\bm{X}Y].
\end{align*}
The matrix $E[\bm{X}\bm{X}^\top]$ can be partitioned into four blocks, as
\begin{align*}
    E[\bm{X}\bm{X}^\top] = \begin{pmatrix}
        1 & E[A] & E[\bm{W}^\top] \\
        E[A] & E[A^2] & E[A\bm{W}^\top] \\
        E[\bm{W}] & E[\bm{W}A] & E[\bm{W}\bm{W}^\top]
    \end{pmatrix} 
    = \begin{pmatrix}
        \bm{S} & \bm{T}^\top \\
        \bm{T} & \bm{U}
    \end{pmatrix}
\end{align*}
where $\bm{S} = 
\begin{pmatrix}
    1 & E[A] \\
    E[A] & E[A^2]
\end{pmatrix}, \bm{T} = (E[\bm{W}], E[\bm{W}A]),$ and  
$\bm{U} = E[\bm{W}\bm{W}^\top]$. If we assume that $\bm{V} = \bm{U} - \bm{T}\bm{S}^{-1}\bm{T}^\top$ is regular, then $E[\bm{X}\bm{X}^\top]$ is also regular, and its inverse matrix is
\begin{align*}
    \{ E[\bm{X}\bm{X}^\top] \}^{-1} = 
    \begin{pmatrix}
        \bm{S}^{-1} + \bm{S}^{-1}\bm{T}^\top\bm{V}^{-1}\bm{T}\bm{S}^{-1} & -\bm{S}^{-1}\bm{T}^\top\bm{V}^{-1} \\
        -\bm{V}^{-1}\bm{T}\bm{S}^{-1} & \bm{V}^{-1}
    \end{pmatrix}.
\end{align*}
Because $A$ is binary, we have
\begin{align*}
    \bm{S}^{-1} = 
    \begin{pmatrix}
        1 & E[A] \\
        E[A] & E[A]
    \end{pmatrix}^{-1}
    = \frac{1}{Var(A)} 
    \begin{pmatrix}
        E[A] & -E[A] \\
        -E[A] & 1
    \end{pmatrix}
\end{align*}
and, under assumption (\ref{eq:independent allocation}), 
\begin{align*}
    \bm{V} &= E[\bm{W}\bm{W}^\top] - 
    (E[\bm{W}], E[\bm{W}]E[A]) \bm{S}^{-1}
    \begin{pmatrix}
        E[\bm{W}^\top] \\
        E[A]E[\bm{W}^\top]
    \end{pmatrix} \\
    &= E[\bm{W}\bm{W}^\top] - \frac{1}{Var(A)} (E[A]-E[A]^2) E[\bm{W}]E[\bm{W}^\top] \\
    &= E[\bm{W}\bm{W}^\top] - E[\bm{W}]E[\bm{W}^\top] \\
    &= Var(\bm{W}).
\end{align*}
Thus, if we assume that $Var(\bm{W})$ is regular, it follows that
\begin{align*}
    -\bm{V}^{-1}\bm{T}\bm{S}^{-1} 
    &= -Var(\bm{W})^{-1} (E[\bm{W}], E[\bm{W}]E[A]) \bm{S}^{-1} \\
    &= -Var(\bm{W})^{-1} \frac{1}{Var(A)} 
    \begin{pmatrix}
        (E[A]-E[A]^2) E[\bm{W}] & 0
    \end{pmatrix} \\
    &= (-Var(\bm{W})^{-1}E[\bm{W}], 0).
\end{align*}
Therefore, we have
\begin{align*}
    \underline{\bm{\beta}}_{\bm{W}} 
    &= (-\bm{V}^{-1}\bm{T}\bm{S}^{-1}, \bm{V}^{-1}) E[\bm{X}Y] \\
    &= (-Var(\bm{W})^{-1}E[\bm{W}], 0, Var(\bm{W})^{-1})
    \begin{pmatrix}
        E[Y] \\
        E[AY] \\
        E[\bm{W}Y]
    \end{pmatrix} \\
    &= Var(\bm{W})^{-1} (E[\bm{W}Y]-E[\bm{W}]E[Y]) \\
    &= Var(\bm{W})^{-1}Cov(\bm{W}, Y).
\end{align*}
This coincides with the probability limit of $\widehat{\bm{\gamma}}_{\bm{W}}$, which is the OLS estimator of the ANCOVA model without the treatment term $E[Y|\bm{W}] = \gamma_0 + \bm{\gamma}_{\bm{W}}^\top\bm{W}$.

\newpage

\section{Baseline characteristics of selected participants from the ACTG 175 trial}

\begin{table}[H]
\centering
\caption{Baseline characteristics of selected participants from the ACTG 175 trial.}
\begin{tabular}{lccc}
  \hline
   & ZDV & ZDV+ddl & Total \\
  Characteristic & $(N=73)$ & $(N=73)$ & $(N=146)$ \\ 
  \hline
  Male sex -- No. (\%) & 62 (84.9) & 60 (82.2) & 122 (83.6)  \\ 
  Age (year) -- Mean (SD) & 35.6 (8.4) & 34.4 (8.9) & 35.0 (8.6) \\ 
  Race -- No. (\%) & & & \\
  \quad White & 58 (79.5)  & 58 (79.5) & 116 (79.5)  \\ 
  \quad Non-white & 15 (20.5) & 15 (20.5) & 30 (20.5)  \\ 
  Weight (kg) -- Mean (SD) & 77.7 (11.7) & 76.0 (16.1) & 76.9 (14.1) \\ 
  Risk factors -- No. (\%) & & & \\
  \quad Homosexual activity & 50 (68.5) & 53 (72.6) & 103 (70.5) \\
  \quad Intravenous-drug use & 11 (15.1) & 8 (11.0) & 19 (13.0) \\ 
  \quad Hemophilia & 4 (5.5)  & 5 (6.8) & 9 (6.2)  \\ 
  Karnofsky score -- Mean (SD) & 95.2 (5.6) & 96.2 (4.9) & 95.7 (5.2) \\ 
  Symptomatic HIV infection -- No. (\%) & 11 (15.1)  & 12 (16.4)  & 23 (15.8)  \\ 
  CD4 cell count (cells/mm$^3$) -- Mean (SD) & 351.9 (130.5) & 355.4 (178.5) & 353.7 (155.8) \\ 
  CD8 cell count (cells/mm$^3$) -- Mean (SD) & 942.2 (430.6) & 1005.5 (574.0) & 973.8 (506.6) \\
  Antiretroviral history -- No. (\%) & 41 (56.2) & 43 (58.9) & 84 (57.5) \\ 
   \hline
\end{tabular}
\label{tab:baseline}
\end{table}

\begin{table}[H]
\centering
\caption{Sample correlations between CD4 count at 20 $\pm$ 5 weeks (outcome) and baseline covariates in selected participants of the ACTG 175 trial.}
\begin{tabular}{lr}
  \hline
  Characteristic & Correlation \\
  \hline
  Sex & $-0.129$ \\
  Age & $0.069$ \\
  Race & $-0.062$ \\
  Weight & $-0.117$ \\
  Homosexual activity & $-0.052$ \\
  Intravenous-drug use & $0.078$ \\
  Hemophilia & $-0.072$ \\
  Karnofsky score & $-0.086$ \\
  Symptomatic HIV infection & $-0.268$ \\
  CD4 cell count & $0.516$ \\
  CD8 cell count & $0.097$ \\
  Antiretroviral history & $-0.186$ \\ 
   \hline
\end{tabular}
\label{tab:corr}
\end{table}

\newpage

\section{Simulation}
\subsection{Data generating mechanism for scenario (4)}\label{suppsec:scenario4}
Latent variables corresponding to the continuous baseline covariates (CD4 count (CD4), CD8 count (CD8), age (AGE), weight (WT), and Karnofsky score (KARN)) and binary baseline covariates (hemophilia (HEMO), homosexual activity (HOMO), history of intravenous drug use (DRUG), race (RACE), sex (SEX), antiretroviral history (HIST), and symptomatic status (SYMP)) were generated from a multivariate normal distribution with mean zero and correlation matrix
\begin{align}
    \Sigma=
    \left(
    \begin{array}{cccccccccccc}
        1 & 0.227 & -0.043 & 0.014 & 0.047 & -0.028 & -0.024 & -0.016 & -0.045 & -0.078 & -0.148 & -0.275 \\
         & 1 & 0.058 & 0.061 & 0.011 & -0.078 & 0.133 & -0.052 & -0.016 & 0.142 & -0.029 & 0.031 \\
         & & 1 & 0.188 & -0.115 & -0.462 & 0.300 & 0.135 & -0.080 & 0.100 & 0.080 & 0.046 \\
         & & & 1 & 0.031 & -0.147 & 0.217 & 0.015 & -0.074 & 0.295 & -0.122 & 0.010 \\
         & & & & 1 & 0.178 & -0.099 & -0.097 & 0.075 & -0.007 & -0.116 & -0.145 \\
         & & & & & 1 & -0.523 & -0.300 & -0.167 & 0.186 & 0.225 & -0.185 \\
         & & & & & & 1 & -0.340 & -0.441 & 0.673 & -0.066 &0.234 \\
         & & & & & & & 1 & 0.220 & -0.402 & 0.015 & 0.023 \\
         & & & & & & & & 1 & -0.537 & -0.118 & -0.130 \\
         & & & & & & & & & 1 & 0.020 & 0.127 \\
         & & & & & & & & & & 1 & 0.147 \\
         & & & & & & & & & & & 1
    \end{array}
    \right).\notag
\end{align}
The continuous baseline covariates were generated by shifting and scaling the latent variables using the following means and standard deviations:
\begin{align}
    E\left[
    \begin{array}{c}
        \mathrm{CD4} \\
        \mathrm{CD8} \\
        \mathrm{AGE} \\
        \mathrm{WT} \\
        \mathrm{KARN}
    \end{array}
    \right] =
    \left(
    \begin{array}{c}
        350.986 \\
        995.702 \\
        35.228 \\
        75.472 \\
        95.484
    \end{array}
    \right),\quad
    SD\left(
    \begin{array}{c}
        \mathrm{CD4} \\
        \mathrm{CD8} \\
        \mathrm{AGE} \\
        \mathrm{WT} \\
        \mathrm{KARN}
    \end{array}
    \right) = 
    \left(
    \begin{array}{c}
        122.303 \\
        481.423 \\
        8.773 \\
        13.429 \\
        5.872
    \end{array}
    \right).\notag
\end{align}
Moreover, the binary baseline covariates were generated by thresholding the latent variables to achieve the following marginal probabilities:
\begin{align}
    P\left(\left(
    \begin{array}{c}
        \text{HEMO} \\
        \text{HOMO} \\
        \text{DRUG} \\
        \text{RACE} \\
        \text{SEX} \\
        \text{HIST} \\
        \text{SYMP}
    \end{array}\right) = 1\right)
    = \left(
    \begin{array}{c}
        0.081 \\
        0.652 \\
        0.129 \\
        0.279 \\
        0.822 \\
        0.586 \\
        0.176
    \end{array}\right)\notag.
\end{align}
The treatment assignment indicator $A$ was generated independently generated from a Bernoulli distribution with $P(A=1) = 0.5$, and the outcome $Y$ was generated from a normal distribution with the following conditional means, given the baseline covariates and $A$:
\begin{align*}
    E[Y|A=0,\bm{W}] &=
    126.771 + 0.719(\text{CD4}) - 0.022(\text{CD8}) - 0.432(\text{AGE}) - 0.455(\text{WT}) + 0.607(\text{KARN}) \\
    &- 58.747(\text{HEMO}) - 19.672(\text{HOMO}) - 10.567(\text{DRUG}) - 5.818(\text{RACE}) + 18.900(\text{SEX}) \\
    &- 41.816(\text{HIST}) - 11.039(\text{SYMP}), \\
    E[Y|A=1,\bm{W}] &= 
    151.054 + 0.591(\text{CD4}) - 0.019(\text{CD8}) + 1.497(\text{AGE}) - 0.446(\text{WT}) + 1.174(\text{KARN}) \\
    &- 36.124(\text{HEMO}) - 34.286(\text{HOMO}) + 16.484(\text{DRUG}) - 41.874(\text{RACE}) + 4.947(\text{SEX}) \\
    &- 52.895(\text{HIST}) - 28.643(\text{SYMP}).
\end{align*}
The conditional variances of the outcome were $Var(Y|A=1,\bm{W}) = (115.510)^2$ and $Var(Y|A=0,\bm{W}) = (108.619)^2$. 

\newpage

\subsection{Additional results}

\begin{table}[H]
    \centering
    \caption{Results of proposed method applied in scenario (1). Monte Carlo bias (Bias), empirical standard error (EmpSE), Monte Carlo power (Power), 95\% coverage (Cov.), and summary of recalculated sample size ($N_\mathrm{Avg}$: mean, $N_\mathrm{Med}$: median, $N_\mathrm{Min}$: minimum, and $N_\mathrm{Max}$: maximum).}
    \begin{tabular}{cccrrrrrrrr}
    \hline
    $\Delta$ & $\sigma_{W_1W_2}$ & $\sigma_{Y\bm{W}}$ & Bias & EmpSE & Power & Cov. & $N_\mathrm{Avg}$ & $N_\mathrm{Med}$ & $N_\mathrm{Min}$ & $N_\mathrm{Max}$ \\
    \hline
    0.3 & 0.25 & 0.25 & -0.001 & 0.104 & 0.812 & 0.951 & 332 & 330 & 246 & 478 \\ 
    0.3 & 0.25 & 0.50 & -0.001 & 0.104 & 0.809 & 0.950 & 223 & 222 & 175 & 334 \\ 
    0.3 & 0.25 & 0.75 & -0.000 & 0.048 & 1.000 & 0.950 & 175 & 175 & 175 & 175 \\ 
    0.3 & 0.50 & 0.25 & -0.001 & 0.105 & 0.811 & 0.950 & 338 & 334 & 252 & 504 \\ 
    0.3 & 0.50 & 0.50 & -0.002 & 0.105 & 0.807 & 0.949 & 247 & 246 & 175 & 372 \\ 
    0.3 & 0.50 & 0.75 & 0.000 & 0.076 & 0.974 & 0.949 & 175 & 175 & 175 & 175 \\ 
    0.3 & 0.75 & 0.25 & -0.001 & 0.104 & 0.814 & 0.952 & 342 & 338 & 254 & 508 \\ 
    0.3 & 0.75 & 0.50 & -0.001 & 0.104 & 0.810 & 0.951 & 264 & 264 & 175 & 388 \\ 
    0.3 & 0.75 & 0.75 & -0.000 & 0.091 & 0.905 & 0.950 & 175 & 175 & 175 & 206 \\ 
    0.4 & 0.25 & 0.25 & -0.003 & 0.137 & 0.814 & 0.951 & 192 & 190 & 112 & 314 \\ 
    0.4 & 0.25 & 0.50 & -0.004 & 0.137 & 0.809 & 0.950 & 129 & 128 & 99 & 226 \\ 
    0.4 & 0.25 & 0.75 & -0.000 & 0.065 & 1.000 & 0.949 & 99 & 99 & 99 & 99 \\ 
    0.4 & 0.50 & 0.25 & -0.002 & 0.137 & 0.817 & 0.951 & 195 & 192 & 128 & 326 \\ 
    0.4 & 0.50 & 0.50 & -0.004 & 0.138 & 0.811 & 0.949 & 143 & 142 & 99 & 236 \\ 
    0.4 & 0.50 & 0.75 & 0.000 & 0.102 & 0.972 & 0.950 & 99 & 99 & 99 & 106 \\ 
    0.4 & 0.75 & 0.25 & -0.002 & 0.138 & 0.815 & 0.950 & 198 & 196 & 134 & 336 \\ 
    0.4 & 0.75 & 0.50 & -0.003 & 0.138 & 0.809 & 0.951 & 153 & 152 & 99 & 252 \\ 
    0.4 & 0.75 & 0.75 & -0.000 & 0.121 & 0.901 & 0.951 & 99 & 99 & 99 & 148 \\ 
    0.5 & 0.25 & 0.25 & -0.005 & 0.172 & 0.811 & 0.950 & 125 & 124 & 63 & 232 \\ 
    0.5 & 0.25 & 0.50 & -0.007 & 0.171 & 0.805 & 0.951 & 85 & 84 & 63 & 164 \\ 
    0.5 & 0.25 & 0.75 & 0.000 & 0.082 & 1.000 & 0.950 & 63 & 63 & 63 & 63 \\ 
    0.5 & 0.50 & 0.25 & -0.004 & 0.171 & 0.812 & 0.951 & 127 & 126 & 64 & 252 \\ 
    0.5 & 0.50 & 0.50 & -0.008 & 0.172 & 0.802 & 0.949 & 94 & 94 & 63 & 190 \\ 
    0.5 & 0.50 & 0.75 & -0.001 & 0.129 & 0.967 & 0.950 & 63 & 63 & 63 & 80 \\ 
    0.5 & 0.75 & 0.25 & -0.003 & 0.172 & 0.813 & 0.950 & 129 & 126 & 63 & 252 \\ 
    0.5 & 0.75 & 0.50 & -0.006 & 0.171 & 0.808 & 0.950 & 100 & 100 & 63 & 196 \\ 
    0.5 & 0.75 & 0.75 & -0.001 & 0.153 & 0.891 & 0.951 & 64 & 63 & 63 & 106 \\ 
    0.6 & 0.25 & 0.25 & -0.006 & 0.203 & 0.815 & 0.950 & 90 & 88 & 44 & 176 \\ 
    0.6 & 0.25 & 0.50 & -0.013 & 0.202 & 0.802 & 0.949 & 62 & 60 & 44 & 128 \\ 
    0.6 & 0.25 & 0.75 & -0.000 & 0.099 & 1.000 & 0.950 & 44 & 44 & 44 & 44 \\ 
    0.6 & 0.50 & 0.25 & -0.008 & 0.203 & 0.812 & 0.951 & 91 & 90 & 44 & 176 \\ 
    0.6 & 0.50 & 0.50 & -0.011 & 0.203 & 0.803 & 0.950 & 68 & 68 & 44 & 140 \\ 
    0.6 & 0.50 & 0.75 & -0.000 & 0.156 & 0.962 & 0.950 & 44 & 44 & 44 & 68 \\ 
    0.6 & 0.75 & 0.25 & -0.008 & 0.204 & 0.812 & 0.949 & 92 & 90 & 44 & 176 \\ 
    0.6 & 0.75 & 0.50 & -0.010 & 0.204 & 0.806 & 0.950 & 72 & 72 & 44 & 158 \\ 
    0.6 & 0.75 & 0.75 & -0.004 & 0.181 & 0.883 & 0.951 & 45 & 44 & 44 & 90 \\ 
    0.7 & 0.25 & 0.25 & -0.012 & 0.234 & 0.817 & 0.950 & 69 & 68 & 33 & 132 \\ 
    0.7 & 0.25 & 0.50 & -0.017 & 0.232 & 0.805 & 0.951 & 48 & 48 & 33 & 114 \\ 
    0.7 & 0.25 & 0.75 & 0.000 & 0.116 & 1.000 & 0.950 & 33 & 33 & 33 & 40 \\ 
    0.7 & 0.50 & 0.25 & -0.011 & 0.234 & 0.816 & 0.949 & 70 & 70 & 33 & 132 \\ 
    0.7 & 0.50 & 0.50 & -0.018 & 0.234 & 0.804 & 0.950 & 52 & 52 & 33 & 128 \\ 
    0.7 & 0.50 & 0.75 & -0.001 & 0.181 & 0.958 & 0.950 & 33 & 33 & 33 & 64 \\ 
    0.7 & 0.75 & 0.25 & -0.010 & 0.234 & 0.819 & 0.950 & 71 & 70 & 33 & 132 \\ 
    0.7 & 0.75 & 0.50 & -0.016 & 0.234 & 0.808 & 0.949 & 56 & 56 & 33 & 130 \\ 
    0.7 & 0.75 & 0.75 & -0.008 & 0.207 & 0.881 & 0.952 & 35 & 33 & 33 & 76 \\ 
    \hline
    \end{tabular}
    \label{tab:scenario1}
\end{table}

\begin{table}[H]
    \centering
    \caption{Results of proposed method applied in scenario (2). Monte Carlo bias (Bias), empirical standard error (EmpSE), Monte Carlo power (Power), 95\% coverage (Cov.), and summary of recalculated sample size ($N_\mathrm{Avg}$: mean, $N_\mathrm{Med}$: median, $N_\mathrm{Min}$: minimum, and $N_\mathrm{Max}$: maximum).}
    \begin{tabular}{cccrrrrrrrr}
    \hline
    $\Delta$ & $\mu_{W_2|W_1}^{(2)}$ & $\beta^{(2)}$ & Bias & EmpSE & Power & Cov. & $N_\mathrm{Avg}$ & $N_\mathrm{Med}$ & $N_\mathrm{Min}$ & $N_\mathrm{Max}$ \\
    \hline
    0.3 & 0.5 & 0.1 & -0.001 & 0.104 & 0.813 & 0.951 & 363 & 354 & 290 & 528 \\ 
    0.3 & 0.5 & 0.3 & -0.002 & 0.105 & 0.810 & 0.949 & 317 & 316 & 222 & 496 \\ 
    0.3 & 0.5 & 0.5 & -0.002 & 0.105 & 0.808 & 0.949 & 226 & 226 & 175 & 332 \\ 
    0.3 & 1.0 & 0.1 & -0.000 & 0.105 & 0.813 & 0.950 & 361 & 354 & 284 & 526 \\ 
    0.3 & 1.0 & 0.3 & -0.001 & 0.104 & 0.813 & 0.950 & 303 & 302 & 210 & 434 \\ 
    0.3 & 1.0 & 0.5 & -0.002 & 0.104 & 0.813 & 0.950 & 189 & 186 & 175 & 272 \\ 
    0.3 & 1.5 & 0.1 & -0.001 & 0.105 & 0.811 & 0.949 & 359 & 352 & 286 & 530 \\ 
    0.3 & 1.5 & 0.3 & -0.001 & 0.104 & 0.809 & 0.951 & 284 & 284 & 175 & 422 \\ 
    0.3 & 1.5 & 0.5 & 0.000 & 0.091 & 0.904 & 0.950 & 175 & 175 & 175 & 206 \\ 
    0.4 & 0.5 & 0.1 & -0.002 & 0.137 & 0.816 & 0.951 & 209 & 204 & 150 & 354 \\ 
    0.4 & 0.5 & 0.3 & -0.003 & 0.137 & 0.814 & 0.950 & 183 & 182 & 112 & 312 \\ 
    0.4 & 0.5 & 0.5 & -0.004 & 0.137 & 0.809 & 0.951 & 131 & 130 & 99 & 226 \\ 
    0.4 & 1.0 & 0.1 & -0.002 & 0.138 & 0.813 & 0.950 & 209 & 204 & 146 & 366 \\ 
    0.4 & 1.0 & 0.3 & -0.003 & 0.137 & 0.815 & 0.951 & 175 & 174 & 99 & 292 \\ 
    0.4 & 1.0 & 0.5 & -0.004 & 0.136 & 0.816 & 0.951 & 110 & 108 & 99 & 186 \\ 
    0.4 & 1.5 & 0.1 & -0.002 & 0.138 & 0.816 & 0.951 & 207 & 202 & 142 & 342 \\ 
    0.4 & 1.5 & 0.3 & -0.003 & 0.138 & 0.812 & 0.949 & 164 & 164 & 99 & 304 \\ 
    0.4 & 1.5 & 0.5 & -0.000 & 0.122 & 0.899 & 0.949 & 99 & 99 & 99 & 138 \\ 
    0.5 & 0.5 & 0.1 & -0.004 & 0.171 & 0.814 & 0.951 & 136 & 132 & 78 & 252 \\ 
    0.5 & 0.5 & 0.3 & -0.004 & 0.172 & 0.812 & 0.949 & 120 & 118 & 64 & 234 \\ 
    0.5 & 0.5 & 0.5 & -0.009 & 0.171 & 0.802 & 0.949 & 86 & 86 & 63 & 168 \\ 
    0.5 & 1.0 & 0.1 & -0.004 & 0.172 & 0.812 & 0.950 & 136 & 132 & 84 & 250 \\ 
    0.5 & 1.0 & 0.3 & -0.006 & 0.172 & 0.808 & 0.949 & 114 & 114 & 63 & 210 \\ 
    0.5 & 1.0 & 0.5 & -0.008 & 0.168 & 0.812 & 0.949 & 73 & 70 & 63 & 134 \\ 
    0.5 & 1.5 & 0.1 & -0.003 & 0.172 & 0.816 & 0.950 & 135 & 132 & 76 & 252 \\ 
    0.5 & 1.5 & 0.3 & -0.006 & 0.172 & 0.807 & 0.950 & 107 & 106 & 63 & 202 \\ 
    0.5 & 1.5 & 0.5 & -0.002 & 0.152 & 0.889 & 0.951 & 64 & 63 & 63 & 106 \\ 
    0.6 & 0.5 & 0.1 & -0.006 & 0.203 & 0.817 & 0.951 & 98 & 94 & 48 & 176 \\ 
    0.6 & 0.5 & 0.3 & -0.008 & 0.204 & 0.811 & 0.950 & 86 & 86 & 44 & 176 \\ 
    0.6 & 0.5 & 0.5 & -0.012 & 0.203 & 0.801 & 0.950 & 62 & 62 & 44 & 130 \\ 
    0.6 & 1.0 & 0.1 & -0.007 & 0.203 & 0.815 & 0.950 & 97 & 94 & 52 & 176 \\ 
    0.6 & 1.0 & 0.3 & -0.009 & 0.204 & 0.809 & 0.950 & 82 & 82 & 44 & 176 \\ 
    0.6 & 1.0 & 0.5 & -0.011 & 0.198 & 0.812 & 0.951 & 53 & 52 & 44 & 112 \\ 
    0.6 & 1.5 & 0.1 & -0.007 & 0.204 & 0.814 & 0.950 & 97 & 94 & 44 & 176 \\ 
    0.6 & 1.5 & 0.3 & -0.010 & 0.204 & 0.806 & 0.950 & 77 & 76 & 44 & 164 \\ 
    0.6 & 1.5 & 0.5 & -0.003 & 0.182 & 0.881 & 0.951 & 45 & 44 & 44 & 94 \\ 
    0.7 & 0.5 & 0.1 & -0.008 & 0.234 & 0.820 & 0.951 & 75 & 72 & 33 & 132 \\ 
    0.7 & 0.5 & 0.3 & -0.012 & 0.234 & 0.815 & 0.950 & 66 & 66 & 33 & 132 \\ 
    0.7 & 0.5 & 0.5 & -0.020 & 0.233 & 0.802 & 0.950 & 48 & 48 & 33 & 118 \\ 
    0.7 & 1.0 & 0.1 & -0.010 & 0.233 & 0.820 & 0.951 & 75 & 72 & 33 & 132 \\ 
    0.7 & 1.0 & 0.3 & -0.014 & 0.234 & 0.812 & 0.950 & 63 & 62 & 33 & 132 \\ 
    0.7 & 1.0 & 0.5 & -0.018 & 0.227 & 0.811 & 0.951 & 41 & 40 & 33 & 102 \\ 
    0.7 & 1.5 & 0.1 & -0.008 & 0.234 & 0.820 & 0.949 & 74 & 72 & 33 & 132 \\ 
    0.7 & 1.5 & 0.3 & -0.015 & 0.235 & 0.809 & 0.950 & 60 & 60 & 33 & 132 \\ 
    0.7 & 1.5 & 0.5 & -0.008 & 0.208 & 0.878 & 0.952 & 35 & 33 & 33 & 72 \\
    \hline
    \end{tabular}
    \label{tab:scenario2}
\end{table}

\begin{table}[H]
    \centering
    \caption{Results of proposed method applied in scenario (3). Monte Carlo bias (Bias), empirical standard error (EmpSE), Monte Carlo power (Power), 95\% coverage (Cov.), and summary of recalculated sample size ($N_\mathrm{Avg}$: mean, $N_\mathrm{Med}$: median, $N_\mathrm{Min}$: minimum, and $N_\mathrm{Max}$: maximum).}
    \begin{tabular}{cccrrrrrrrr}
    \hline
    $\Delta$ & $\mu_{W_2|W_1}^{(3)}$ & $\beta^{(3)}$ & Bias & EmpSE & Power & Cov. & $N_\mathrm{Avg}$ & $N_\mathrm{Med}$ & $N_\mathrm{Min}$ & $N_\mathrm{Max}$ \\
    \hline
    0.3 & 0.2 & 0.1 & -0.001 & 0.105 & 0.813 & 0.948 & 366 & 356 & 306 & 532 \\ 
    0.3 & 0.2 & 0.3 & -0.001 & 0.104 & 0.812 & 0.951 & 349 & 344 & 258 & 510 \\ 
    0.3 & 0.2 & 0.5 & -0.001 & 0.104 & 0.812 & 0.950 & 314 & 312 & 222 & 472 \\ 
    0.3 & 0.4 & 0.1 & -0.000 & 0.104 & 0.814 & 0.950 & 366 & 356 & 302 & 562 \\ 
    0.3 & 0.4 & 0.3 & -0.001 & 0.105 & 0.811 & 0.949 & 345 & 342 & 260 & 494 \\ 
    0.3 & 0.4 & 0.5 & -0.001 & 0.104 & 0.810 & 0.950 & 304 & 302 & 212 & 488 \\ 
    0.3 & 0.6 & 0.1 & -0.001 & 0.105 & 0.812 & 0.948 & 365 & 356 & 302 & 550 \\ 
    0.3 & 0.6 & 0.3 & -0.001 & 0.105 & 0.812 & 0.950 & 342 & 338 & 258 & 494 \\ 
    0.3 & 0.6 & 0.5 & -0.001 & 0.105 & 0.809 & 0.949 & 295 & 294 & 198 & 434 \\ 
    0.4 & 0.2 & 0.1 & -0.003 & 0.138 & 0.813 & 0.951 & 211 & 204 & 154 & 358 \\ 
    0.4 & 0.2 & 0.3 & -0.002 & 0.138 & 0.817 & 0.950 & 201 & 198 & 136 & 338 \\ 
    0.4 & 0.2 & 0.5 & -0.003 & 0.138 & 0.813 & 0.950 & 181 & 180 & 100 & 322 \\ 
    0.4 & 0.4 & 0.1 & -0.002 & 0.138 & 0.816 & 0.950 & 211 & 204 & 150 & 380 \\ 
    0.4 & 0.4 & 0.3 & -0.003 & 0.138 & 0.815 & 0.951 & 199 & 196 & 130 & 340 \\ 
    0.4 & 0.4 & 0.5 & -0.003 & 0.138 & 0.814 & 0.950 & 176 & 176 & 106 & 290 \\ 
    0.4 & 0.6 & 0.1 & -0.002 & 0.137 & 0.818 & 0.951 & 211 & 204 & 146 & 360 \\ 
    0.4 & 0.6 & 0.3 & -0.002 & 0.138 & 0.816 & 0.950 & 198 & 196 & 122 & 348 \\ 
    0.4 & 0.6 & 0.5 & -0.003 & 0.138 & 0.811 & 0.950 & 171 & 170 & 100 & 280 \\ 
    0.5 & 0.2 & 0.1 & -0.003 & 0.171 & 0.815 & 0.950 & 138 & 132 & 88 & 252 \\ 
    0.5 & 0.2 & 0.3 & -0.005 & 0.171 & 0.812 & 0.950 & 131 & 128 & 72 & 252 \\ 
    0.5 & 0.2 & 0.5 & -0.004 & 0.171 & 0.812 & 0.951 & 118 & 118 & 63 & 218 \\ 
    0.5 & 0.4 & 0.1 & -0.003 & 0.171 & 0.815 & 0.951 & 137 & 132 & 84 & 252 \\ 
    0.5 & 0.4 & 0.3 & -0.004 & 0.172 & 0.813 & 0.950 & 130 & 128 & 70 & 238 \\ 
    0.5 & 0.4 & 0.5 & -0.004 & 0.172 & 0.811 & 0.951 & 115 & 114 & 63 & 226 \\ 
    0.5 & 0.6 & 0.1 & -0.004 & 0.171 & 0.813 & 0.951 & 137 & 132 & 86 & 252 \\ 
    0.5 & 0.6 & 0.3 & -0.005 & 0.171 & 0.811 & 0.951 & 129 & 126 & 70 & 248 \\ 
    0.5 & 0.6 & 0.5 & -0.005 & 0.171 & 0.809 & 0.951 & 112 & 110 & 63 & 210 \\ 
    0.6 & 0.2 & 0.1 & -0.006 & 0.204 & 0.816 & 0.949 & 99 & 94 & 50 & 176 \\ 
    0.6 & 0.2 & 0.3 & -0.005 & 0.203 & 0.817 & 0.951 & 94 & 92 & 44 & 176 \\ 
    0.6 & 0.2 & 0.5 & -0.008 & 0.204 & 0.810 & 0.951 & 85 & 84 & 44 & 168 \\ 
    0.6 & 0.4 & 0.1 & -0.006 & 0.204 & 0.815 & 0.949 & 98 & 94 & 48 & 176 \\ 
    0.6 & 0.4 & 0.3 & -0.006 & 0.204 & 0.815 & 0.950 & 93 & 92 & 44 & 176 \\ 
    0.6 & 0.4 & 0.5 & -0.009 & 0.204 & 0.810 & 0.951 & 83 & 82 & 44 & 166 \\ 
    0.6 & 0.6 & 0.1 & -0.006 & 0.203 & 0.815 & 0.952 & 98 & 94 & 48 & 176 \\ 
    0.6 & 0.6 & 0.3 & -0.007 & 0.204 & 0.813 & 0.950 & 92 & 90 & 44 & 176 \\ 
    0.6 & 0.6 & 0.5 & -0.010 & 0.204 & 0.808 & 0.950 & 80 & 80 & 44 & 176 \\ 
    0.7 & 0.2 & 0.1 & -0.009 & 0.234 & 0.821 & 0.951 & 76 & 72 & 33 & 132 \\ 
    0.7 & 0.2 & 0.3 & -0.011 & 0.233 & 0.819 & 0.951 & 72 & 70 & 33 & 132 \\ 
    0.7 & 0.2 & 0.5 & -0.013 & 0.234 & 0.814 & 0.951 & 65 & 64 & 33 & 132 \\ 
    0.7 & 0.4 & 0.1 & -0.010 & 0.234 & 0.822 & 0.950 & 76 & 72 & 33 & 132 \\ 
    0.7 & 0.4 & 0.3 & -0.010 & 0.234 & 0.819 & 0.950 & 72 & 70 & 33 & 132 \\ 
    0.7 & 0.4 & 0.5 & -0.013 & 0.235 & 0.812 & 0.949 & 64 & 64 & 33 & 132 \\ 
    0.7 & 0.6 & 0.1 & -0.008 & 0.234 & 0.822 & 0.950 & 76 & 72 & 33 & 132 \\ 
    0.7 & 0.6 & 0.3 & -0.011 & 0.235 & 0.817 & 0.949 & 71 & 70 & 33 & 132 \\ 
    0.7 & 0.6 & 0.5 & -0.014 & 0.235 & 0.810 & 0.950 & 62 & 62 & 33 & 132 \\
    \hline
    \end{tabular}
    \label{tab:scenario3}
\end{table}

\newpage

\begin{table}[H]
\centering
\caption{Results of proposed method applied in scenario (4) with misspecified within-arm variance of the outcome at the design stage. For underestimation, it was set to $0.9\sigma_Y^2$ (Under) and for overestimation, it was set to $1.1\sigma_Y^2$ (Over). Monte Carlo bias (Bias), Monte Carlo percent bias (\%Bias), empirical standard error (EmpSE), Monte Carlo power (Power), 95\% coverage (Cov.), and summary of recalculated sample sizes ($N_\mathrm{Avg}$: mean, $N_\mathrm{Med}$: median, $N_\mathrm{Min}$: minimum, and $N_\mathrm{Max}$: maximum) for the ANCOVA estimator adjusting for (1) CD4 count only, (2) antiretroviral history only, (3) CD4 count and antiretroviral history, (4) all continuous baseline covariates, (5) all binary baseline covariates, and (6) all baseline covariates.}
\begin{tabular}{llrrrrrrrrr}
  \hline
 & & Bias & \%Bias & EmpSE & Power & Cov. & $N_\mathrm{Avg}$ & $N_\mathrm{Med}$ & $N_\mathrm{Min}$ & $N_\mathrm{Max}$ \\ 
  \hline
Under & (1) & $-1.097$ & $-1.561$ & $24.427$ & $0.791$ & $0.950$ & $94$ & $94$ & $59$ & $194$ \\ 
   & (2) & $-0.715$ & $-1.017$ & $24.563$ & $0.797$ & $0.949$ & $132$ & $126$ & $70$ & $236$ \\ 
   & (3) & $-1.197$ & $-1.703$ & $24.555$ & $0.785$ & $0.950$ & $91$ & $90$ & $59$ & $188$ \\ 
   & (4) & $-1.277$ & $-1.816$ & $25.085$ & $0.771$ & $0.950$ & $93$ & $92$ & $59$ & $202$ \\ 
   & (5) & $-1.036$ & $-1.473$ & $25.295$ & $0.770$ & $0.950$ & $126$ & $122$ & $60$ & $236$ \\ 
   & (6) & $-1.984$ & $-2.823$ & $26.413$ & $0.724$ & $0.949$ & $87$ & $86$ & $59$ & $186$ \\ 
  \hline
   & (1) & $-0.696$ & $-0.990$ & $23.249$ & $0.837$ & $0.950$ & $104$ & $104$ & $73$ & $184$ \\ 
   & (2) & $-0.325$ & $-0.463$ & $23.218$ & $0.845$ & $0.950$ & $146$ & $146$ & $86$ & $244$ \\ 
   & (3) & $-0.789$ & $-1.122$ & $23.359$ & $0.833$ & $0.949$ & $101$ & $100$ & $73$ & $188$ \\ 
   & (4) & $-0.836$ & $-1.190$ & $23.763$ & $0.820$ & $0.949$ & $103$ & $102$ & $73$ & $184$ \\ 
   & (5) & $-0.607$ & $-0.863$ & $23.802$ & $0.823$ & $0.950$ & $139$ & $138$ & $73$ & $246$ \\ 
   & (6) & $-1.368$ & $-1.947$ & $24.739$ & $0.781$ & $0.949$ & $97$ & $96$ & $73$ & $170$ \\ 
   \hline
\end{tabular}
\label{tab:under over}
\end{table}

\newpage

\subsection{Additional simulations and results}
\subsubsection{Simulation results with the misspecified working ANCOVA model}\label{suppsubsubsec:misspecified}
To evaluate the robustness of the proposed method under model misspecification, we conducted simulations in which the working ANCOVA model was misspecified. Two baseline covariates $W_1$ and $W_2$ were independently generated from standard normal distributions, and treatment $A$ was randomly assigned in a 1:1 ratio independently of $(W_1,W_2)$. The outcome $Y$ was generated from a normal distribution with conditional mean
\begin{align}
    E[Y\mid A,W_1,W_2] = \Delta A+0.2\times W_1^2+0.3\times\exp(W_2),\notag
\end{align}
within arm variance $\sigma_Y^2=1$, and five effect sizes, $\Delta=\{0.3,0.4,0.5,0.6,0.7\}$. 

We assumed the working ANCOVA model
\begin{align}
    E[Y|A,W_1,W_2]=\beta_0+\beta_A A+\beta_{W_1}W_1+\beta_{W_2}W_2,\notag
\end{align}
which omitted the quadratic term $W_1^2$ and the nonlinear term $\exp(W_2)$, and was therefore misspecified. The effect size and within-arm variance were otherwise correctly specified. We considered a one-sided test with type I error rate $\alpha=0.025$ and power $1-\beta=0.8$. The interim analysis was conducted at $\tau=0.5$, and the upper bound of the available sample size was set to twice the required sample size for the unadjusted estimator $(m=2)$. We conducted $100,000$ Monte Carlo simulations for each value of $\Delta$. 

The results are presented in Supplementary Table~\ref{tab:misspecified}. The results were broadly similar to those obtained the correctly specified working ANCOVA model. In particular, the proposed method appeared slightly conservative, which we conjecture arises from the use of the $\min()$ term in the denominator of the proposed formula. Nevertheless, the proposed method reduced the final sample size relative to the unadjusted estimator for all effect sizes, for which $N_\mathrm{unadj}=\{350,198,126,88,66\}$.

\begin{table}[hbtp]
\centering
\caption{Results of the proposed method with the misspecified working ANCOVA model. Monte Carlo bias (Bias), empirical standard error (EmpSE), Monte Carlo power (Power), 95\% coverage (Cov.), and summary of recalculated sample size ($N_\mathrm{Avg}$: mean, $N_\mathrm{Med}$: median, $N_\mathrm{Min}$: minimum, and $N_\mathrm{Max}$: maximum).}
\begin{tabular}{lrrrrrrrr}
  \hline
  $\Delta$ & Bias & EmpSE & Power & Cov. & $N_\mathrm{Avg}$ & $N_\mathrm{Med}$ & $N_\mathrm{Min}$ & $N_\mathrm{Max}$ \\ 
  \hline
  $0.3$ & $-0.001$ & $0.102$ & $0.831$ & $0.951$ & $290$ & $280$ & $188$ & $700$ \\ 
  $0.4$ & $-0.004$ & $0.134$ & $0.833$ & $0.950$ & $167$ & $162$ & $99$ & $396$ \\ 
  $0.5$ & $-0.004$ & $0.167$ & $0.833$ & $0.950$ & $109$ & $106$ & $63$ & $252$ \\ 
  $0.6$ & $-0.008$ & $0.197$ & $0.835$ & $0.951$ & $78$ & $76$ & $44$ & $176$ \\ 
  $0.7$ & $-0.014$ & $0.229$ & $0.834$ & $0.950$ & $59$ & $58$ & $33$ & $132$ \\ 
  \hline
\end{tabular}
\label{tab:misspecified}
\end{table}

\newpage

\subsubsection{Simulation results with stratified randomization}\label{suppsubsubsec:car}
We conducted simulations to evaluate the performance of the proposed method under stratified randomization. Two dichotomous baseline covariates $W_1$ and $W_2$ were generated as
\begin{align}
    W_1\sim Ber(0.5),\quad W_2\mid W_1\sim Ber(0.5+0.4\times(W_1-0.5)).\notag
\end{align}
Treatment $A$ was assigned using stratified randomization with a block size of 4 based on the stratified factors of $(W_1,W_2)$. The outcome $Y$ was generated from a normal distribution with conditional mean
\begin{align}
    E[Y\mid A,W_1,W_2] = \Delta A+0.3\times W_1 + 0.3\times W_2,\notag
\end{align}
within-arm variance $\sigma_Y^2=1$, and five effect sizes, $\Delta=\{0.3,0.4,0.5,0.6,0.7\}$. We assumed the working ANCOVA model was correctly specified and that the effect size and within-arm variance were correctly specified at the design stage. We considered a one-sided test with type I error rate $\alpha=0.025$ and power $1-\beta=0.8$. The interim analysis was conducted at $\tau=0.5$ and the upper bound of the available sample size was set to twice the required sample size for the unadjusted estimator $(m=2)$. We conducted $100,000$ Monte Carlo simulations for each value of $\Delta$.

The results are presented in Supplementary Table~\ref{tab:car}. The empirical results under stratified randomization were broadly similar to those under complete randomization (scenario (3), $\mu_{W_2\mid W_1}^{(3)}=0.4, \beta^{(3)}=0.3$), although empirical power was slightly higher under stratified randomization.

\begin{table}[hbtp]
\centering
\caption{Results of the proposed method under stratified randomization. Monte Carlo bias (Bias), empirical standard error (EmpSE), Monte Carlo power (Power), 95\% coverage (Cov.), and summary of recalculated sample size ($N_\mathrm{Avg}$: mean, $N_\mathrm{Med}$: median, $N_\mathrm{Min}$: minimum, and $N_\mathrm{Max}$: maximum).}
\begin{tabular}{lrrrrrrrr}
  \hline
  $\Delta$ & Bias & EmpSE & Power & Cov. & $N_\mathrm{Avg}$ & $N_\mathrm{Med}$ & $N_\mathrm{Min}$ & $N_\mathrm{Max}$ \\ 
  \hline
  $0.3$ & $-0.001$ & $0.104$ & $0.815$ & $0.950$ & $345$ & $342$ & $266$ & $500$ \\ 
  $0.4$ & $-0.002$ & $0.137$ & $0.820$ & $0.949$ & $200$ & $196$ & $122$ & $342$ \\ 
  $0.5$ & $-0.004$ & $0.169$ & $0.822$ & $0.950$ & $130$ & $128$ & $78$ & $252$ \\ 
  $0.6$ & $-0.007$ & $0.200$ & $0.826$ & $0.951$ & $94$ & $92$ & $44$ & $176$ \\ 
  $0.7$ & $-0.009$ & $0.227$ & $0.837$ & $0.950$ & $72$ & $70$ & $33$ & $132$ \\ 
  \hline
\end{tabular}
\label{tab:car}
\end{table}

\newpage

\subsubsection{Simulation results for scenario (4) in large samples}\label{suppsubsubsec:large-sample-size}
We conducted simulations to evaluate the performance of the recalculation methods in large samples. The data-generating mechanism was identical to that of scenario (4), except that the intercepts of the outcome models were modified so that the resulting marginal treatment effect was $10.303$. This yielded a substantially larger required sample size for the unadjusted estimator $(N_\mathrm{unadj}=6460)$.\\
We applied both the proposed recalculation method and the simple recalculation method using ANCOVA models adjusting for (1) CD4 count only, (2) antiretroviral history only, (3) CD4 count and antiretroviral history, (4) all continuous baseline covariates, (5) all binary baseline covariates, and (6) all baseline covariates. $1,000$ Monte Carlo runs were conducted for each adjustment strategy.

The results are presented in Supplementary Table~\ref{tab:scenario4 large}. Both the simple recalculation method and the proposed method achieved the nominal power of 80\% even when (6) all baseline covariates were adjusted for. These findings suggest that the reduction in power observed in Table~\ref{tab:scenario4} may be attributable to finite-sample limitations of the asymptotic approximation.

\begin{table}[hbtp]
\centering
\caption{Results for scenario (4) in large samples from 1,000 Monte Carlo runs. Monte Carlo bias (Bias), Monte Carlo percent bias (\%Bias), empirical standard error (EmpSE), Monte Carlo power (Power), 95\% coverage (Cov.), and summary of recalculated sample size ($N_\mathrm{Avg}$: mean, $N_\mathrm{Med}$: median, $N_\mathrm{Min}$: minimum, and $N_\mathrm{Max}$: maximum) for ANCOVA estimator, adjusting for (1) CD4 count only, (2) antiretroviral history only, (3) CD4 count and antiretroviral history, (4) all continuous baseline covariates, (5) all binary baseline covariates, and (6) all baseline covariates. "Simple" and "Prop." indicate the simple recalculation method and the proposed method, respectively.}
\begin{tabular}{lllrrrrrrrrr}
  \hline
Estimator & & Method & Bias & \%Bias & EmpSE & Power & Cov. & $N_\mathrm{Avg}$ & $N_\mathrm{Med}$ & $N_\mathrm{Min}$ & $N_\mathrm{Max}$ \\ 
  \hline
Unadjusted &  &  & $-0.056$ & $-0.546$ & $3.692$ & $0.792$ & $0.942$ & \multicolumn{4}{c}{$6460$} \\ 
  ANCOVA & (1) & Simple & $-0.002$ & $-0.019$ & $3.699$ & $0.796$ & $0.947$ & $4451$ & $4453$ & $4044$ & $4814$ \\ 
   &  & Prop. & $0.012$ & $0.119$ & $3.710$ & $0.798$ & $0.945$ & $4492$ & $4488$ & $4224$ & $4808$ \\ 
  ANCOVA & (2) & Simple & $0.002$ & $0.020$ & $3.677$ & $0.799$ & $0.950$ & $6150$ & $6150$ & $5616$ & $6696$ \\ 
   &  & Prop. & $-0.004$ & $-0.036$ & $3.650$ & $0.800$ & $0.952$ & $6209$ & $6188$ & $6008$ & $6690$ \\ 
  ANCOVA & (3) & Simple & $0.004$ & $0.043$ & $3.679$ & $0.802$ & $0.948$ & $4292$ & $4292$ & $3924$ & $4642$ \\ 
   &  & Prop. & $0.016$ & $0.157$ & $3.663$ & $0.802$ & $0.946$ & $4331$ & $4325$ & $4022$ & $4636$ \\ 
  ANCOVA & (4) & Simple & $-0.018$ & $-0.175$ & $3.703$ & $0.790$ & $0.942$ & $4399$ & $4400$ & $4002$ & $4730$ \\ 
   &  & Prop. & $-0.004$ & $-0.043$ & $3.711$ & $0.793$ & $0.942$ & $4440$ & $4436$ & $4188$ & $4724$ \\ 
  ANCOVA & (5) & Simple & $0.092$ & $0.893$ & $3.692$ & $0.804$ & $0.954$ & $5879$ & $5880$ & $5414$ & $6394$ \\ 
   &  & Prop. & $0.058$ & $0.567$ & $3.681$ & $0.799$ & $0.950$ & $5936$ & $5924$ & $5682$ & $6388$ \\ 
  ANCOVA & (6) & Simple & $0.037$ & $0.364$ & $3.701$ & $0.794$ & $0.950$ & $4144$ & $4144$ & $3798$ & $4464$ \\ 
   &  & Prop. & $0.027$ & $0.260$ & $3.653$ & $0.799$ & $0.952$ & $4182$ & $4178$ & $3918$ & $4458$ \\ 
   \hline
\end{tabular}
\label{tab:scenario4 large}
\end{table}

\newpage

\section{Code}\label{suppsec:code}
\begin{lstlisting}[caption=R code to implement the proposed method]
n.ancova.recal <- function(outcome, cov, data, delta, sigma, alpha=0.025, power=0.8) {
  
  z.a <- qnorm(1 - alpha)
  z.b <- qnorm(power)
  
  fo <- as.formula(paste0(outcome, "~", paste(cov, collapse="+")))
  
  model <- lm(fo, data=data)
  res <- summary(model)
  
  num <- res$sigma^2 - delta^2/4
  den <- min(sigma, var(data[, outcome]) - delta^2/4)
  
  n.unadj <- ceiling(2^2 * ((z.a + z.b)^2) * sigma / (delta^2))
  n.unadj <- n.unadj + (n.unadj %% 2)
  
  n.ancova <- ceiling(n.unadj * num / den + z.a^2/2)
  n.ancova <- n.ancova + (n.ancova %% 2)
  
  return(n.ancova)
}
\end{lstlisting}

\begin{lstlisting}[caption=SAS macro to implement the proposed method]
%macro n_ancova_recal(outcome, cov, data, delta, sigma, alpha=0.025, power=0.8);
ods output ANOVA=res_ancova;
proc reg data=&data.;
    model &outcome. = &cov.;
quit;

ods output Summary=res_unadj;
proc means data=&data.;
    var &outcome.;
run;

data res_ancova;
    set res_ancova;
    num = SS / DF - &delta.**2 / 4;
    if Source = "Error" then output;
    keep num;
run;

data res_unadj;
    set res_unadj;
    den = min(&sigma., &outcome._Var);
    keep den;
run;

data res;
    merge res_ancova res_unadj;
    z_a = quantile("normal", 1-&alpha.);
    z_b = quantile("normal", &power.);
    n_unadj = ceil(2**2 * ((z_a+z_b)**2) * &sigma. / (&delta.**2));
    n_unadj = n_unadj + mod(n_unadj, 2);
    n_ancova = ceil(n_unadj * num / den + z_a**2/2);
    n_ancova = n_ancova + mod(n_ancova, 2);
run;

title "Recalculated sample size for ANCOVA";
proc print data=res;
    var n_ancova;
run;
%mend;
\end{lstlisting}

\end{document}